\documentclass[fleqn,usenatbib]{mnras}

\usepackage{newtxtext,newtxmath}

\usepackage[T1]{fontenc}

\usepackage{bm}
\DeclareRobustCommand{\VAN}[3]{#2}
\let\VANthebibliography\thebibliography
\def\thebibliography{\DeclareRobustCommand{\VAN}[3]{##3}\VANthebibliography}

\usepackage{graphicx}	
\usepackage{amsmath}	
\usepackage{subfig}
\usepackage{subcaption}
\usepackage{hyperref} 
\usepackage{xcolor}
\usepackage[normalem]{ulem} 

\title[TUNA for Radio Astronomy]{Mapping Diffuse Radio Sources Using TUNA: A Transformer-Based Deep Learning Approach}

\author[N. Sanvitale et al.]{N. Sanvitale$^1$\thanks{E-mail: nicoletta.sanvitale@inaf.it}, C. Gheller$^1$, F. Vazza$^2$, A. Bonafede$^{2}$, V. Cuciti$^2$, E. De Rubeis$^2$, F. Govoni$^3$, M. Murgia$^3$, V. Vacca$^3$\\
$^{1}$ Istituto di Radio Astronomia, INAF, Via P. Gobetti 101, 40129 Bologna, Italy\\
$^{2}$ Dipartimento di Fisica e Astronomia, Universit\'{a} di Bologna, Via P. Gobetti 92/3, 40129 Bologna, Italy\\
$^3$ Osservatorio Astronomico di Cagliari, INAF, Via della Scienza 5, 09047 Cagliari, Italy
}

\date{Accepted XXX. Received YYY; in original form ZZZ}

\pubyear{\the\year{}}

\begin{document}
\label{firstpage}
\pagerange{\pageref{firstpage}--\pageref{lastpage}}
\maketitle

\begin{abstract}
Vision Transformers are used via a customized TransUNet architecture, which is a hybrid model combining Transformers into a U-Net backbone, to achieve precise, automated, and fast segmentation of radio astronomy data affected by calibration and imaging artifacts, addressing the identification of faint, diffuse radio sources. Trained on mock radio observations from numerical simulations, the network is applied to the LOFAR Two-meter Sky Survey data. It is then evaluated on key use cases, specifically megahalos and bridges between galaxy clusters, to assess its performance in targeting sources at different resolutions and at the sensitivity limits of the telescope. The network is capable of detecting low surface brightness radio emission without manual source subtraction or re-imaging. The results demonstrate its groundbreaking capability to identify sources that typically require reprocessing at resolutions 4–6 times lower than that of the input image, accurately capturing their morphology and ensuring detection completeness. This approach represents a significant advancement in accelerating discovery within the large datasets generated by next-generation radio telescopes.
\end{abstract}




\section{Introduction}
\label{sec:introduction}

Radio observations of galaxy clusters provide invaluable insights into the properties of the non-thermal components of the intra-cluster medium, revealing the presence of cosmic-ray electrons and magnetic fields across volumes of several cubic megaparsecs. These components may extend beyond the cluster's virial radius, potentially permeating the cosmic web \citep[see e.g.][]{ry03, pf06, 2008Sci...320..909R, 2025arXiv250119041V}. 
Radio emission, primarily found in massive, dynamically active galaxy clusters, is traditionally classified as giant radio halos, mini halos, and radio relics based on their morphology, origin, and cluster location \citep[for reviews][]{feretti2012clusters, brunetti2014cosmic, van2019diffuse, wittor2023cosmic}. 
Modern low frequency radio-interferometric observations, characterised by unprecedented  sensitivity, revealed synchrotron-emitting bridges connecting pairs of massive clusters in a pre-merger phase \citep[see e.g.][]{govoni2019radio, botteon2020giant, 2024A&A...691A..99P}, possibly formed through turbulence in the intra-cluster region \citep{brunetti20}, providing the first evidence of cosmic structures that trace the large-scale structure of the universe. These observations have also allowed the discovery of a new class of large diffuse radio sources, outside the typical radius of ``classical" radio halos ($\geq 0.4 R_{\rm 500}$ out to $R_{\rm 500}$, where $R_{\rm 500}$ is the spherical radius enclosing a total matter overdensity 500 times greater than the cosmic mean density at the corresponding redshift), referred to as ``megahalos''  \citep{cuciti2022galaxy}. Bridges and megahalos are predicted to be rare compared to classical radio halos and relics, since they might originate during specific stages of the cluster evolution and in presence of particular merger configurations \citep{brunetti20}. The detection of the few occurrences present in current and forthcoming radio surveys is thus of paramount importance.

The common feature of these observations is that they target extended sources, challenging to detect given their low surface brightness emission combined with their large angular scale. Characterizing their morphology and properties requires overcoming the limitations imposed by the constrained observational sensitivity due to factors like radio frequency interference, ionospheric effects, and limited dynamic range. Furthermore, even in an ideal scenario of perfectly calibrated data, detecting extended emission necessitates accurate subtraction of compact sources that dominate the interferometric signal. These sources, if not removed properly, can leave sidelobes or residuals interfering with the detection of low-surface-brightness structures. This typically requires re-imaging the data at lower spatial resolutions, which helps enhance sensitivity to large angular scales. However, this re-imaging process is highly computationally demanding and time consuming and it must be carefully tuned, as it can lead to the suppression of real features, to the amplification of random and confusion noise. 

The methodological challenges are worsened by the volume of the data involved. Such observations result from the processing of huge datasets generated by state-of-the-art radio telescopes such as the Low-Frequency Array (LOFAR)  \citep{van2013lofar}, the Karoo Array Telescope (MeerKAT) \citep{jonas2016meerkat}, the Murchison Widefield Array (MWA) \citep{tingay2013murchison}, and the Australian Square Kilometre Array Pathfinder (ASKAP) \citep{hotan2021australian}, whose unprecedented sensitivity, resolution, frequency coverage, and field of view significantly increase data volume. They are precursor to the hundreds of petabytes of data that the Square Kilometre Array Observatory (SKAO\footnote{https://www.skao.int/}) is expected to generate  annually. 
Efficient numerical solutions that are capable of performing tasks such as radio source detection, segmentation, and flux calculation in a fast, accurate, and fully automated manner are necessary to manage such large amounts of complex data, minimising the necessity for human supervision, which becomes impractical for large datasets, and eliminating the need for reprocessing or re-imaging the data through traditional pipelines.
In this work, we explore the adoption of a customized version of the TransUnet, hereafter referred as {\it TransUnet Network for radio Astronomy (TUNA)} to radio observations. The network is trained using mock radio images generated from numerical simulations, then applied to various observational targets selected from the second data release of the LOFAR Two-metre Sky Survey (LoTSS-DR2) and the Planck Sunyaev–Zel’dovich (PSZ2) cluster catalogue \citep{botteon2022planck}. This catalogue serves as a reference for the statistical analysis of TUNA's capabilities and performance.  We also compare these results with those obtained using a method based on the original U-Net architecture \citep{2024MNRAS.533.3194S}, in order to explore the differences and similarities between the two approaches. The TUNA network is subsequently evaluated on several additional cases, specifically megahalos and bridges \citep[][]{cuciti2022galaxy, govoni2019radio, botteon2020giant}, to assess its performance in targeting sources at the limits of LOFAR sensitivity and in complex radio environments.

\section{Related Work}
\label{sec:related_work}

Machine and Deep Learning (ML/DL)  have the potential to effectively address the requirements of large-scale radio astronomy data processing, as demonstrated by the number of applications in different scientific domains published over the last decade \citep[see e.g.][]{rolnick2022climate_change, vamathevan2019applications, radovic2018machine, 2021MNRAS.506..659C}. In particular, the authors have explored the use of DL in radio astronomy, focusing on Convolutional Neural Networks for automated detection of sources \citep{2018MNRAS.480.3749G}, Denoising Autoencoders for image quality enhancement \citep{2022MNRAS.509..990G}, and Fully Convolutional U-Net architecture (hereafter R-UNet) for semantic segmentation of diffuse radio emission \citep{2024MNRAS.533.3194S}.

The advent of the Transformer architecture \citep{vaswani2017attention} has had a profound impact on various domains of AI, particularly on natural language processing (NLP), representing the foundation for breakthrough Large Language Models, such as BERT \citep{devlin-etal-2019-bert}, GPT \citep{radford2018improving}, Gemini \citep{2023arXiv231211805G} and DeepSeek \citep{2024arXiv241219437D}.  At their core, Transformers use a self-attention mechanism \citep{vaswani2017attention}, which allows the model to focus on different parts of the input to capture key relationships and extract intrinsic features from data. 

Transformers quickly extended to other fields, firstly computer vision. This paper investigates the use of Vision Transformers (ViT) \citep{dosovitskiy2020image} to enhance the effectiveness of machine learning methods for processing radio interferometric data. 
ViTs process images as patches, using self-attention to capture relationships across them. TransUNet \citep{chen2021transunet}, introduced in 2021 for medical image segmentation, is a pioneering hybrid model integrating ViTs into the U-Net architecture \citep{ronneberger2015}, combining the extensive contextual awareness of the former with the localized feature extraction abilities of the latter. This enables the identification of extended targets with large-scale correlations, while accurately capturing small-scale features, making it an ideal solution for the identification and segmentation of diffuse radio sources. 

\section{Methods}
\label{sec:methods}

This section presents an overview of the TUNA network architecture, the training data employed, and its performance on mock observations. 

\subsection{The TUNA Architecture}
\label{sec:TransUnet}

TUNA relies on the TransUnet network \citep{chen2021transunet}. This is a hybrid architecture which integrates  the Vision Transformer module, as introduced by \cite{dosovitskiy2020image}, into the U-Net architecture \citep{ronneberger2015} to address the constraints of the limited receptive field of convolutional operations, which hinder U-Net's ability to capture long-range contextual dependencies.

\begin{figure*}
\begin{center}
	\includegraphics[width=1.5\columnwidth]{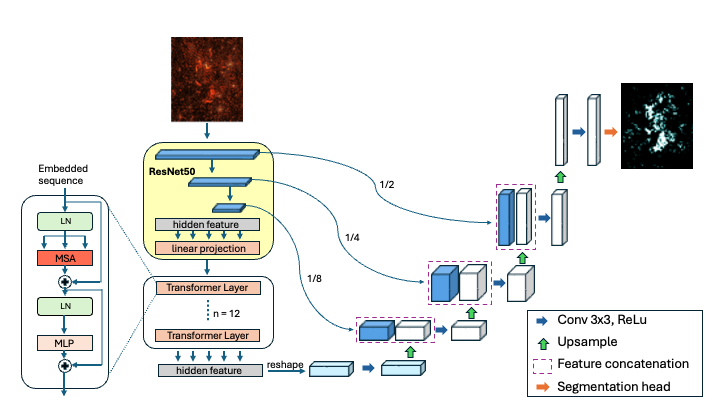}\\ 
    \caption{ Schematic overview of the TransUNet architecture }
    \label{fig:transunet}
\end{center}
\end{figure*}

Fig. \ref{fig:transunet} shows that TransUnet retains the encoder-decoder structure of U-Net but employing a hybrid CNN-Transformer encoder. In this architecture, CNN, specifically ResNet50 \citep{he2016resnet}, serves as a feature extractor, producing feature maps that encapsulate fine-grained spatial information. If the input image has dimensions \( H \times W \times C \) (height, width, and number of channels), the output feature map dimensions are given by:

\begin{equation}
    H^{\prime} = \frac{H}{P}\ ; \quad W^{\prime} = \frac{W}{P}
\end{equation}

where $P$ is the downsampling factor of the CNN backbone ($P=16$ for ResNet50).

Each feature map is then divided into patches of size ($P_s \times P_s$), which are linearly projected into D-dimensional embeddings. The TransUnet architecture assumes $P_s=1$. To encode spatial information, position embeddings are learned and added:
\begin{equation}
    \bm{\mathrm{z}}_0 = [\bm{\mathrm{x}}^1_p \bm{\mathrm{E}}; \, \bm{\mathrm{x}}^2_p \bm{\mathrm{E}}; \cdots; \, \bm{\mathrm{x}}^{N}_p \bm{\mathrm{E}} ] + \bm{\mathrm{E}}_{pos}, \label{eq:embedding} 
\end{equation}
where \{$\bm{\mathrm{x}}^i_p \in \mathbb{R}^{P^2_s \cdot C^\prime}|i=1,..,N\}$ are the flattened 2D patches, $N=\frac{HW}{P^2}$ is the number of image patches (\emph{i.e.}, the input sequence length), $C^\prime$ is the number of feature maps, 
$\bm{\mathrm{E}} \in \mathbb{R}^{(P_s^2 \cdot C^\prime) \times D}$ is the patch embedding projection, and $\bm{\mathrm{E}}_{pos}  \in \mathbb{R}^{N \times D}$ denotes the position embedding.  

The resulting sequence of embedding vectors serves as input to the Transformer encoder.
The Transformer encoder \citep{vaswani2017attention} consists of a stack of n=12 identical layers. Each layer sequentially processes the input, with the output of one layer serving as the input for the next.
Each layer has two blocks. The first is a multi-head self-attention mechanism (MSA), and the second is a simple, positionwise fully connected feed-forward network (MLP). There are residual connections around both the MSA and the MLP blocks. Layer normalization (LN) is applied after each residual connection to stabilize training.
The output of the $l$-th layer can be written as follows:

\begin{eqnarray}
    \bm{\mathrm{z}}^\prime_l &=& \text{MSA}(\text{LN}(\bm{\mathrm{z}}_{l-1})) + \bm{\mathrm{z}}_{l-1},  \label{eq:msa_apply} \\
    \bm{\mathrm{z}}_l &=& \text{MLP}(\text{LN}(\bm{\mathrm{z}}^\prime_{l})) + \bm{\mathrm{z}}^\prime_{l},   \label{eq:mlp_apply} 
\end{eqnarray}
where $\bm{\mathrm{z}}_l$ is the encoded image representation.

The flattened hidden features produced by the Transformer encoder are reshaped back into a 2D spatial structure, making it compatible with the U-Net decoder. The resulting low-resolution feature map is progressively restored to the full resolution H×W to obtain the final segmentation prediction using a stack of bilinear upsampling blocks. Each block sequentially applies a 2× bilinear upsampling (doubles spatial resolution), a 3×3 convolution layer (refines spatial details), and a ReLU activation function (enhances feature learning introducing nonlinearity). The version of TUNA adopted in the paper exploits ResNet-50 and the 12 layer ViT encoder pre-trained on ImageNet21k, the latent dimension D of the embedding space is set to 768.
TUNA employs a hybrid segmentation loss that combines pixel-wise cross-entropy loss ($\mathcal{L}_{CE}$) and Dice Loss($\mathcal{L}_{Dice}$) defined as:
\begin{equation}
    \mathcal{L}_{\text{CE}} =  -\log \left( \frac{\exp(x_y)}{\sum_{j} \exp(x_j)} \right),   
\end{equation}
where $x_j$ is the prediction for class j and $x_y$ is the prediction for the true class $y$, 
\begin{equation}
    \mathcal{L}_{\text{Dice}} = 1 - \frac{2 \sum_{i=1}^{N} y_i \hat{y}_i}{\sum_{i=1}^{N} y_i^2 + \sum_{i=1}^{N} \hat{y}_i^2}, \label{eq:dice_loss}
\end{equation}
where $y_i$ is the ground truth value, $\hat{y}_i$ is the predicted value, and $N$ is the number of pixels. The total loss is:
\begin{equation}
    \mathcal{L} = \lambda_1 \cdot \mathcal{L}_{CE} + \lambda_2 \cdot \mathcal{L}_{Dice},
\end{equation}
where $\lambda_{\rm 1}$ and  $\lambda_{\rm 2}$ are weighting factors set equal to 0.5.
Cross-entropy loss quantifies the difference between predicted class confidence scores and ground truth labels. Dice Loss measures the overlap between predicted and true segmentation masks, emphasizing shape similarity.

The input data for TUNA consists of radio astronomy images, for which a specific data loader has been implemented. Input images are provided in FITS format \citep{1981A&AS...44..363W}, which represents a well-established open standard in astronomy. The data loader implements a FITS reader, based on the AstroPy library \citep{astropy:2013}. Since radio astronomical FITS images can be large and vary significantly in size, an image tiling strategy, based on the approach described in \cite{Sanvitale2022}, has been implemented to enable the handling of images of arbitrary dimensions. Padding is applied as needed to ensure compatibility with the network input requirements. For the training non-overlapping tiles are used. For the inference, overlaps of 50\% or 75\% are supported. Radio astronomical images contains energy fluxes with a large dynamical range. Once read, fluxes are preprocessed to make them usable by the network by calculating their logarithm and normalising the results values. Upon performing inference, the output tiles are converted back to their original units and scale, the image is reconstructed from the tiles and saved in a FITS file. 

\subsection{Training Data}
\label{sec:data}

TUNA was trained on synthetic observations of diffuse radio emission, produced from cosmological simulations. Images have been produced to closely resemble actual LOFAR observations, following the methodology outlined in \cite{2022MNRAS.509..990G} and \cite{2024MNRAS.533.3194S}. Here we summarize the main steps of the procedure.

The cosmological simulations were performed using ideal Magneto-hydrodynamics (MHD) with the Eulerian code {\it Enzo} \citep{enzo14}. The simulation box has a uniform spatial cell resolution of $41.65 \rm ~kpc$ (comoving) and it encompasses a comoving volume of $100^3$Mpc$^3$. A uniform primordial magnetic seed field of $B_0=0.1 {\rm nG}$ (comoving) was initialised everywhere in the box at the start of the simulation ($z=50$) and was later evolved under ideal MHD conditions until $z=0$. The simulation assumed a standard $\Lambda$CDM cosmological model, with density parameters $\Omega_{b} = 0.0478$, $\Omega_{\rm DM} = 0.2602$, $\Omega_\Lambda = 0.692$, and a Hubble parameter of $H_0=67.8 ~\rm km/s/Mpc$.

In post-processing, we calculated the emission at 150 MHz from relativistic electrons accelerated by cosmic shocks at different redshifts, assuming that only shocks can accelerate relativistic particles through diffusive shock acceleration and generate synchrotron radio emission \citep[e.g.][]{2011JApA...32..577B}. The model of synchrotron emission proposed by \cite{hb07} is employed in this context, while we did not account for additional radio emission generated by the reacceleration by turbulence on relativistic electrons, which likely has a key role in the formation  of radio halos, bridges, or megahalos \citep{Brunetti11b,brunetti20,Nishiwaki24}.  However, the morphology and emissivity of diffuse radio sources can loosely resemble even those of radio halos, despite the different particle acceleration mechanisms likely at work in radio halos \citep[e.g.][]{van2019diffuse,Lee24}.

The creation of the training set of synthetic images for TUNA, based on the 3-dimensional emission model from cosmological simulations, followed two main basic steps.
First, we integrated the emission along the line of sight within the same volume at  four different redshift snapshots
(roughly equally spaced from $z=0.02$ to $z=0.15$). 

Next, every single synthetic sky model was generated by progressively stacking maps referred to an increasing redshift. In order to fully cover the $\approx 640 ~\rm Mpc$ distance out to $z=0.15$, we extracted maps multiple times from the same simulated redshifts, applying an appropriate cosmological corrections factor to simulate an increasing redshift of the sources, similar to \cite{2021MNRAS.500.5350V}. As done in \cite{2022MNRAS.509..990G}, we limited our analysis of simulated lightcones up to $z \approx 0.15$ since we do not expect a drastic change in the radio emission including larger redshifts \citep[see also][]{2021PASA...38...47H}.

By applying random rotations to the various redshift slices, over 500 independent lightcones were generated. The images have a resolution of $2000 \times 2000$ pixels, covering a field of view of $1.1^\circ\times1.1^\circ$ each, with a nominal angular resolution (pixel size) of 2 arcsec. These images are referred to as {\it sky images}. 

To generate mock observations, the synthetic visibilities corresponding to a LOFAR HBA observation of 8 hours have been calculated using the ``predict'' mode of the WSClean with the sky images as model images \citep{offringa-wsclean-2014,offringa-wsclean-2017}. At this point, a random Gaussian noise with rms similar to LotSS  noise is added to the visibilities using the noise.py script in the LoSiTo software package\footnote{\url{https://github.com/darafferty/losito/tree/master}}. Finally, imaging is performed using the WSClean software, adopting Uniform weighting and correcting for the primary beam. The use of these parameters resulted in a restoring beam of $5.9\arcsec\times5.1\arcsec$. Deconvolution is carried out using the clean method and the auto-masking option with a 5$\sigma$ threshold. The resulting images are referred to as {\it clean images}. Along with the corresponding sky images, they compose the training and test datasets for TUNA. We will refer to this images as the $6''$ training dataset. From the sky images, the ground truth is generated by assigning a value of 0 to all pixels with surface brightness less than $10^{-8}$ Jy/pixel (Class 0) and a value of 1 to all other pixels (Class 1), creating the binary masks that have to be reproduced by the network. The resulting data set is unbalanced, with Class 1 pixels accounting for only 3\% of the total pixel count on average. The same procedure has been repeated to create a $20''$ training dataset, convolving the sky images with a Gaussian function, to achieve a $20''$ resolution.

\subsection{Network Training, Optimization and Performance Analysis}
\label{sec:optimization}

The following methodology  is used to train and assess TUNA on mock observations. One hundred $2000 \times 2000$ pixel images are divided into $N_T \times N_T$ pixel tiles (generated adding suitable padding) and used for training the network. Ten percent of the data are used as a validation set to assess the network's performance during the training process. Forty additional images are used as a testing dataset to evaluate the final performance of the network. The training has been performed using both $6''$ and $20''$ datasets.

To determine the optimal network configuration, we performed multiple trainings using various hyperparameter settings. More specifically, we have set the learning rate $\mu = [0.01, 0.005, 0.001, 0.0001]$, batch size $B = [12, 24, 48]$ and tile size $N_T = [224, 512]$.  The learning rate controls the step size at which a model updates its weights during training; the batch size defines the number of training samples processed before updating the model’s weights in each iteration; the tile size defines the dimension of a subregion into which the input image is divided for training. 
The network performance has been evaluated on a pixel basis using four metrics: Recall, Precision, Intersection over Union (IoU), and Accuracy, calculated by comparing the inferred mask to the ground truth. The four metrics are defined as follows, computed for the positive class (diffuse emission):

\begin{eqnarray}
    {\rm Recall} &=& \frac{TP}{TP+FN}\\
    {\rm Precision} &=& \frac{TP}{TP+FP}\\
    {\rm IoU} &=& \frac{\rm Intersection}{\rm Union} \nonumber \\ &=& \frac{\sum ({\rm Predicted\ Mask \cap Ground\ Truth\ Mask})}{\sum {\rm Predicted\ Mask \cup Ground\ Truth\ Mask}}\\
    {\rm Accuracy} &=& \frac{TP+TN}{TP+TN+FP+FN}
\end{eqnarray}

where TP (True Positives) and FP (False Positives) represent the number of pixels correctly and incorrectly classified as diffuse emission, while TN (True Negatives) and FN (False Negatives) correspond to pixels correctly and incorrectly classified as non-emitting, respectively. Recall measures the fraction of true sources correctly identified, while Precision quantifies the proportion of predicted sources that are TP. IoU evaluates the overlap between predicted and ground truth masks, and Accuracy represents the overall fraction of correctly classified instances across both diffuse emission and non-emitting regions. 

The performance of the various configurations is summarized in Appendix \ref{sec:hyper}. The best configuration was chosen based on the values of the reported metrics and by visual inspection of the results on the test images. For both $6''$ and $20''$ datasets, it is achieved with the following setup: $\mu = 0.005$, $B = 24$, $T = 512$.

Adopting such ideal setup, the uncertainty associated with the training process was estimated by recalculating the performance metrics through five repeated training runs, following a slight variation of the K-Fold Cross-Validation approach \citep{kohavi1995study}. In each training session, a different subset comprising four-fifths of the total training set is used. The resulting models are then applied for inference on a fixed test set consisting of the same forty images.

\begin{figure}
\begin{center}
	\includegraphics[width=1\columnwidth]{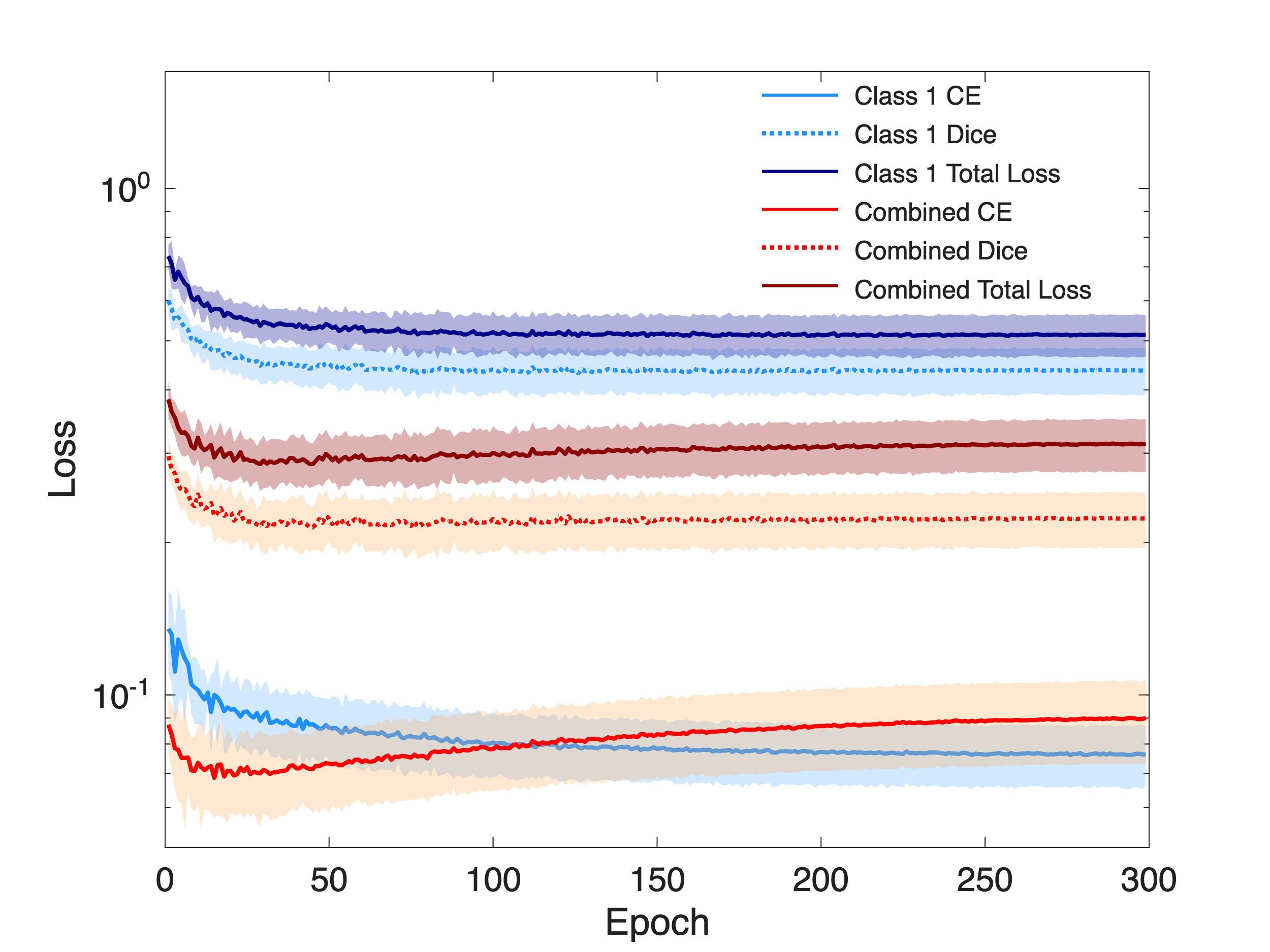}
\end{center}
\caption{TUNA network validation losses as a function of the training epoch for the $6''$ dataset. The blue curves show the Cross Entropy (CE), the Dice and the Total (average of the previous two) losses for Class 1 pixels. The red curves Show the same losses but considering both Classes. Standard deviations are estimated over five different trainings of the network.}
\label{fig:loss}
\end{figure}

The validation loss, that is, the error calculated in the validation dataset after each training epoch (E), is presented in Fig. \ref{fig:loss} for the $6''$ dataset. The figure displays the Cross Entropy (CE) and Dice curves, along with the Total Loss, which is the average of the two losses. The three curves are presented for Class 0 combined with Class 1, as well as for Class 1 only. 

The combined loss curves indicate that training convergence is reached within the first 50 epochs. However, the Class 1 losses keep decreasing up to 200 epochs, indicating that additional training is needed for the network to effectively learn to classify Class 1 pixels. This occurs because the dataset is highly imbalanced with Class 1 constituting approximately $3\%$ of all pixels. 
To address this issue, we applied class weighting in the loss function, assigning a higher weight to the underrepresented Class 1, according to the following equation:
\begin{equation}
\mathcal{L} = \mathcal{L}_{\rm Class0} + w_1 \mathcal{L}_{\rm Class1}    
\end{equation}
where $w_1$ is the weighting factor (the values [1, 2, 10] have been tested) and $\mathcal{L}_{\rm ClassN}$ is the contribution to the loss of Class N. However, this strategy proved ineffective in improving the model's ability to classify Class 1 pixels (see Appendix \ref{sec:hyper}). The only effective solution was to extend the training to 200 epochs. The same approach also proved effective for the $20''$ training set.

Table \ref{tab:metrics_sim} presents the values of the metrics obtained for Class 1 using  the optimal set-up, alongside the corresponding values achieved with the R-UNet model \citep{2024MNRAS.533.3194S}. For the $6''$ model TUNA has a Recall of $0.50\pm 0.01$, whereas R-UNet has a Recall of $0.39\pm 0.01$, indicating an higher effectiveness of the former to identify emitting pixels. Furthermore, TUNA demonstrates superior performance in terms of IoU compared to R-UNet, suggesting that it is more effective also at accurately capturing the morphology of sources. The R-UNet demonstrates slightly higher Precision. This is primarily because R-UNet tends to identify the brightest pixels within the distribution, which are more reliably matched to the ground truth. As a result, its predictions align more accurately with the most prominent structures in the dataset. Finally, the two networks exhibit comparable Accuracy, approaching unity. This reflects the fact that both can precisely identify Class 0 pixels, which predominate the input images. 

At a lower resolution of $20''$, all scores improve as a result of a larger restoring beam, which increases the sensitivity to diffuse emission enhancing the detectability of fainter regions, although at the expenses of the finer details along the source edges.

\begin{figure*}
\begin{center}
    \includegraphics[width=0.8\textwidth]{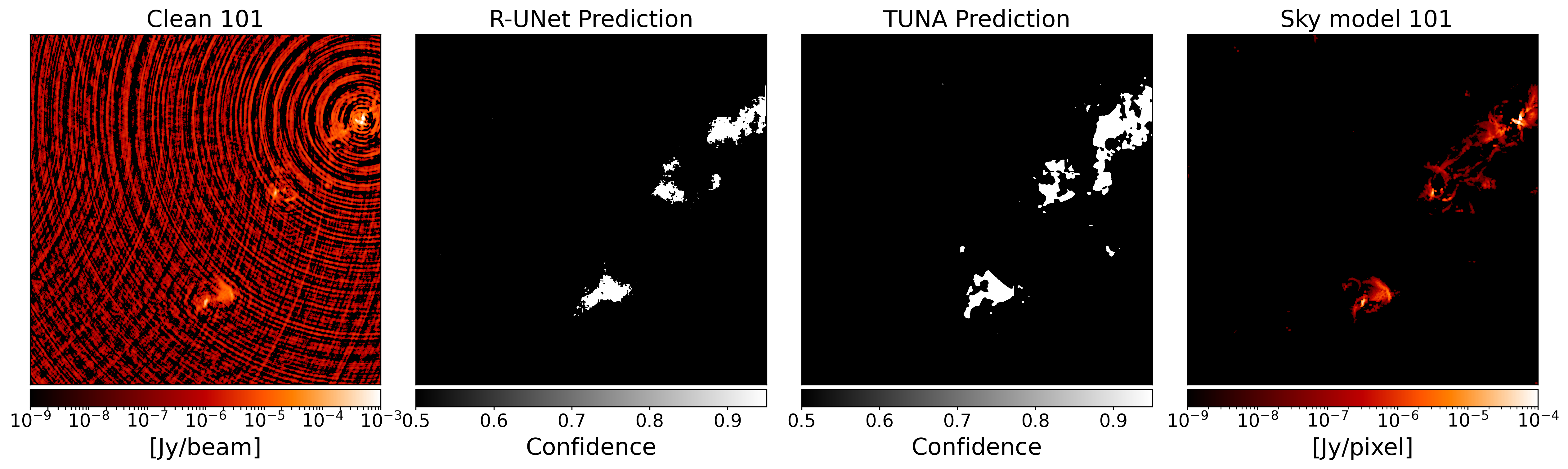}
	\includegraphics[width=0.8\textwidth]{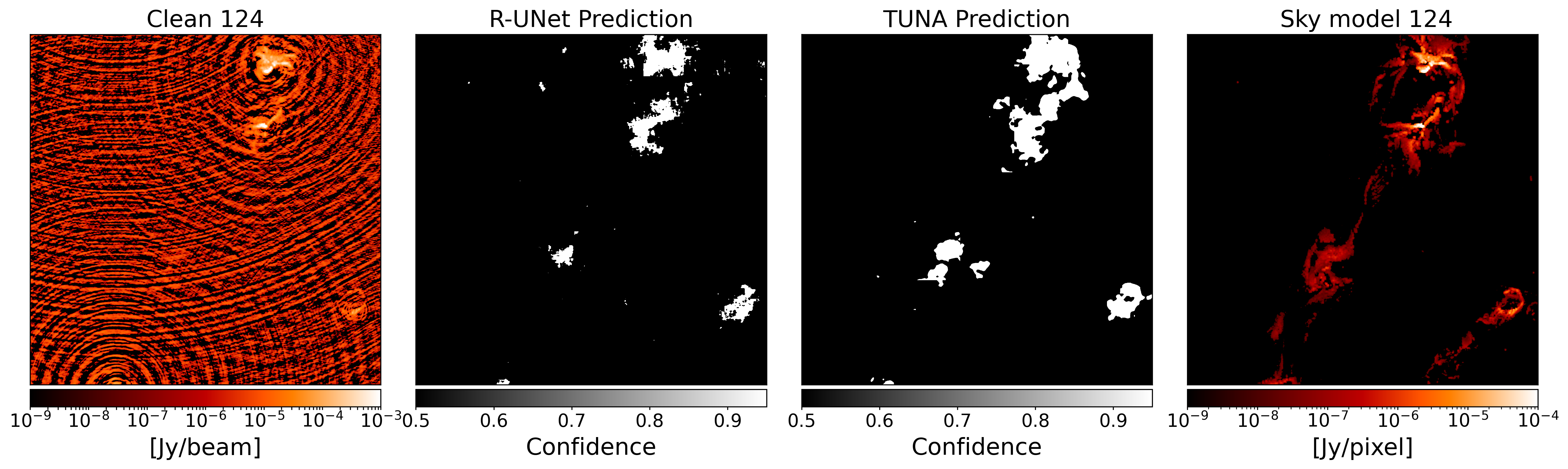}
    \caption{The two rows show examples of Clean images at $6''$ used as input for the network (left panels), sky maps with surface brightness exceeding $10^{-8}$ Jy/pixel (our ground truth, right panels), predictions from TUNA (mid-left panels), and predictions from R-UNet (mid-right panels), all shown at the 0.5 confidence level.}
    \label{fig:comparison110}
\end{center}
\end{figure*}

Fig. \ref{fig:comparison110} presents two examples of mock images used as network input and ground truth for training, as well as predictions given by TUNA and R-UNet at the 0.5 confidence level. Both networks demonstrate the ability to segment the input image even in the presence major artifacts. As also observed above, TUNA is capable of identifying a greater number of pixels at low surface brightness and demonstrates an improved capacity to accurately reproduce the source's actual morphology. 

\begin{figure}
\begin{center}
    \includegraphics[width=0.7\columnwidth]{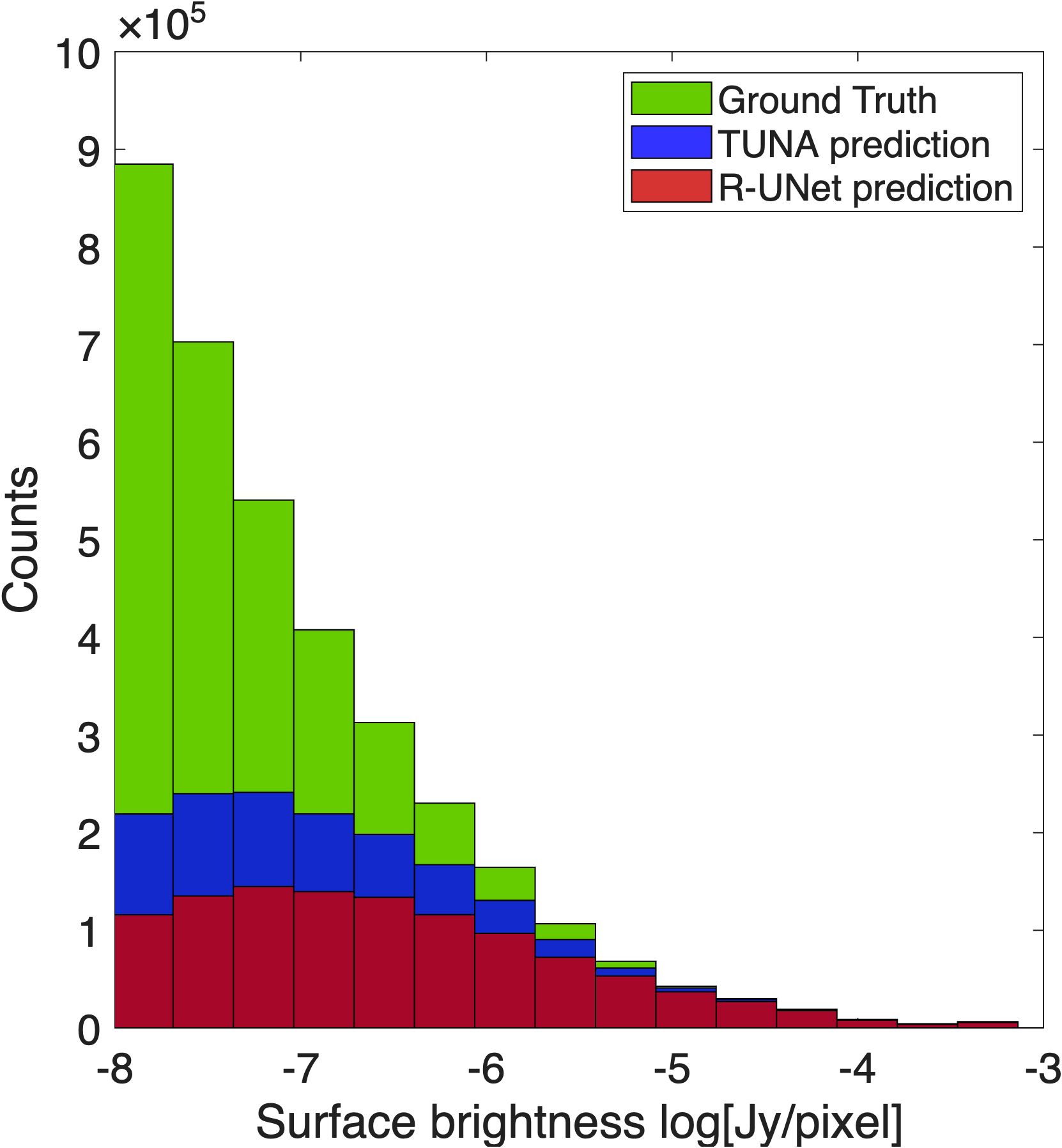}
    \caption{Surface brightness distribution for for the simulated dataset at 6'', considering pixels above $10^{-8}$ Jy/pixel, accounting for the pixels identified by TUNA and R-UNet}
    \label{fig:histomock}
\end{center}
\end{figure}

In Fig. \ref{fig:histomock} the distributions of the surface brightness of the pixels exceeding $10^{-8}$ Jy/pixel along with those selected by the TUNA and by the R-UNet predictions are shown. For surface brightness above $10^{-5}$Jy/pixel, both networks successfully detect all source pixels. At lower values, an increasing fraction of the emitting pixels cannot be detected in the clean image, due to noise and artifacts. However TUNA identifies a larger number of sources than R-UNet and detects a large fraction (more than 50\%) of the emitting pixels down to $10^{-7}$Jy/pixel. Noticeable, at even lower surface brightness, the network can detect a significant fraction of the emitting pixels. 

It is important to mention that all the training and testing were conducted with High Performance Computing resources on the Booster partition of the Leonardo supercomputer\footnote{\url{https://www.hpc.cineca.it/systems/hardware/leonardo/}} at the CINECA Italian Supercomputing centre. The Leonardo system comprises 3456, 32-cores, Intel Xeon Platinum 8358 CPU nodes equipped with 4 NVIDIA Tesla Ampere 100 GPUs. Each CPU node is equipped with 512 GB of DDR4 memory, while the GPU has 64 GB of HBM2 memory. The network can efficiently leverage GPU acceleration thanks to the support provided by the PyTorch framework and can exploit parallel computing for hyperparameters space exploration. The computational performance on a single node of Leonardo is 2.9 sec/image for the inference and 88.5 sec/epoch for the training, using 2000$\times$2000 pixel images.

\begin{table}
	\centering
	\caption{Performance metrics for TUNA and R-UNet calculated for the positive class (diffuse emission). The second and third columns present scores for the simulated data at $6''$ and the last two columns for the simulated data at $20''$.}
	\label{tab:metrics_sim}
	\begin{tabular}{lcccc} 
		\hline
		Metric & TUNA $6''$ & R-UNet $6''$ & TUNA $20''$ & R-UNet $20''$ \\
		\hline
        Recall & $0.50\pm 0.01$ & $0.39\pm 0.01$ & $0.62\pm 0.02$ & $ 0.53\pm 0.01$\\
		Precision & $0.61\pm 0.02$ & $0.64\pm 0.01$ & $0.76\pm 0.02$ & $ 0.73\pm 0.01$\\
        IoU & $0.38\pm 0.01$ & $0.30\pm 0.01$ & $0.52\pm 0.01$ & $ 0.43\pm 0.01$\\
        Accuracy & $0.97\pm 0.01$ & $0.96\pm 0.01$ & $0.97\pm 0.01$ & $ 0.97\pm 0.01$\\
		\hline
	\end{tabular}
\end{table}

\section{Results}
\label{sec:results}

Although trained on mock observations from cosmological simulations, TUNA proves to successfully generalize to real observational data, effectively distinguishing between artifacts and diffuse sources. In order to assess its performance on observational data, the network has been applied to multiple use cases, namely the Planck Catalogue from the LOFAR Two-meter Sky Survey (LoTSS), radio bridges in merging galaxy clusters and mega halos. The observational data are converted to logarithmic scale and normalized between $10^{-8}$ and $10^{-2}$.

\subsection{Planck Catalog from LoTSS-DR2}

The LoTSS-DR2 data is used to evaluate the performance of TUNA on real, observational data. LoTSS-DR2 covers the $27\%$ of the northern sky observed in the range 120-168 MHz (the nominal central frequency is 144 MHz). Each LoTSS pointing is 8 hr long, the typical resolution is $\sim 6''$, and the median root-mean-square (rms) noise is $\sigma \sim 0.083 \; {\rm mJy/beam}$.  The survey includes low-resolution (20$\arcsec$) images with median  sensitivity  $\sigma\sim  0.095\; {\rm mJy/beam}$ \citep{shimwell2022lofar}. 

We used as a benchmark the radio observations of the 309 clusters in PSZ2 that lie within the 5634 deg$^2$ covered by the LoTSS-DR2. This dataset, hereafter referred to as LoTSS-DR2/PSZ2, represents the most comprehensive study to date of diffuse synchrotron emission in the intra-cluster medium. We employed data available from the LoTSS repository\footnote{https://lofar-surveys.org/planck\_dr2.html} produced by \cite{botteon2022planck}. 

\begin{figure*}
    \makebox[\textwidth][l]{ %
        \resizebox{0.87\textwidth}{!}{ 
            \begin{minipage}{1\textwidth}
                \centering
                \subfloat{\raisebox{43ex}{\hspace{0ex}\footnotesize (a)}\includegraphics[width=1.1\textwidth]
                {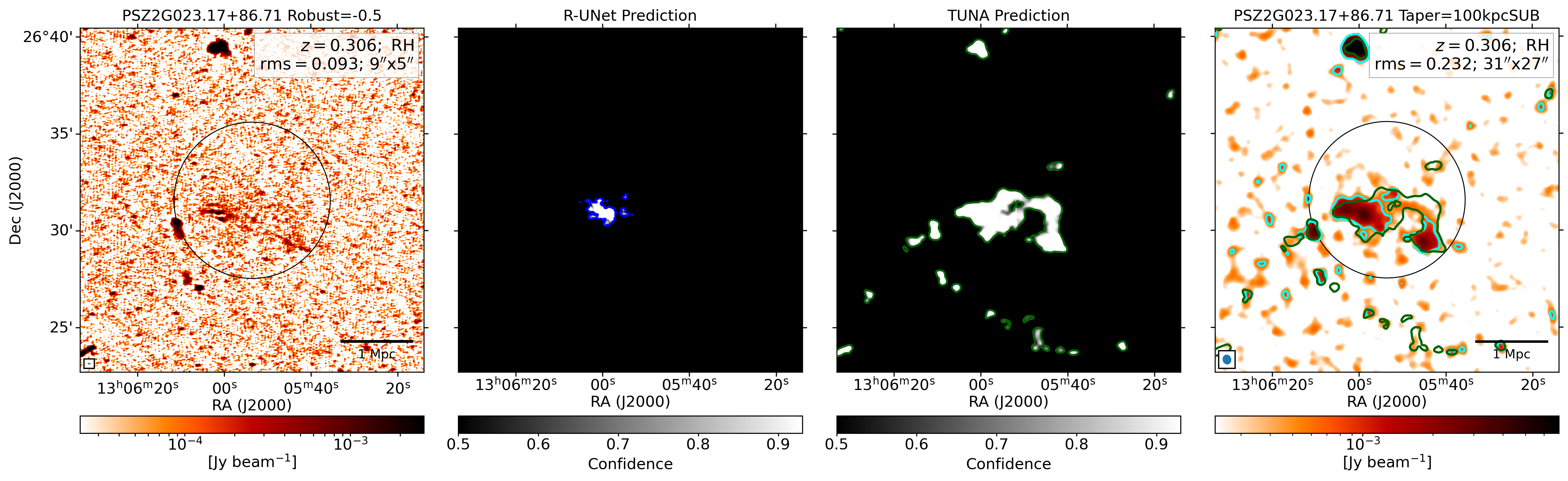}}\\
                \vspace{-1mm} 
                \subfloat{\raisebox{43ex}{\hspace{0ex}\footnotesize (b)}\includegraphics[width=1.1\textwidth]{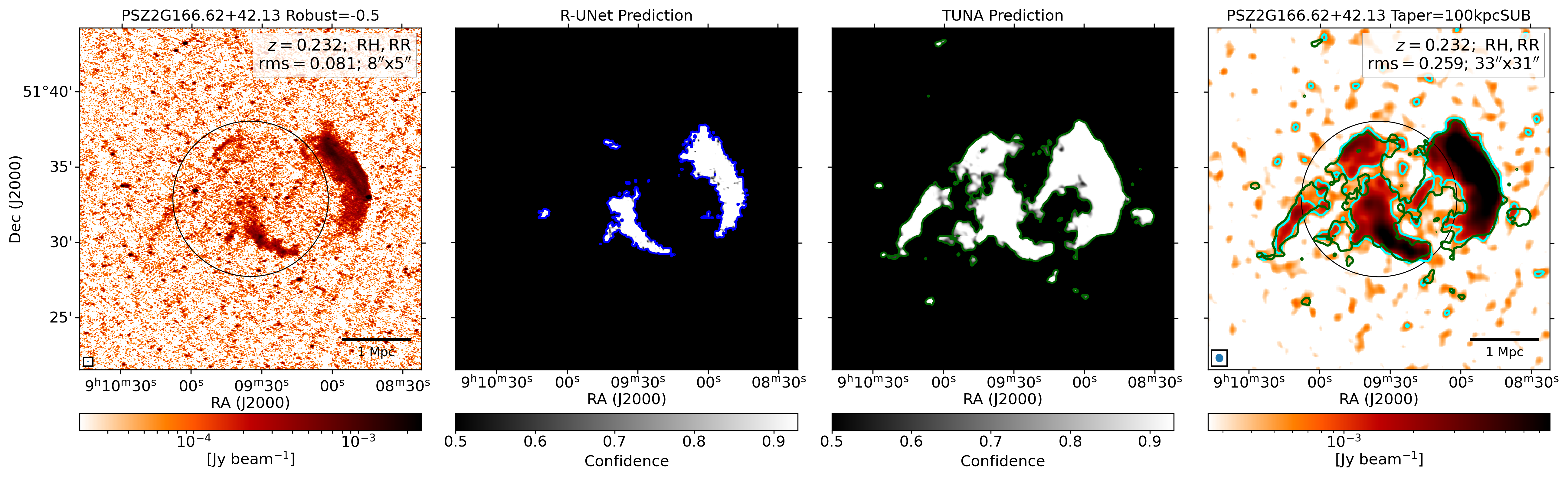}}\\
                \vspace{-1mm} 
                \subfloat{\raisebox{43ex}{\hspace{0ex}\footnotesize (c)}\includegraphics[width=1.1\textwidth]{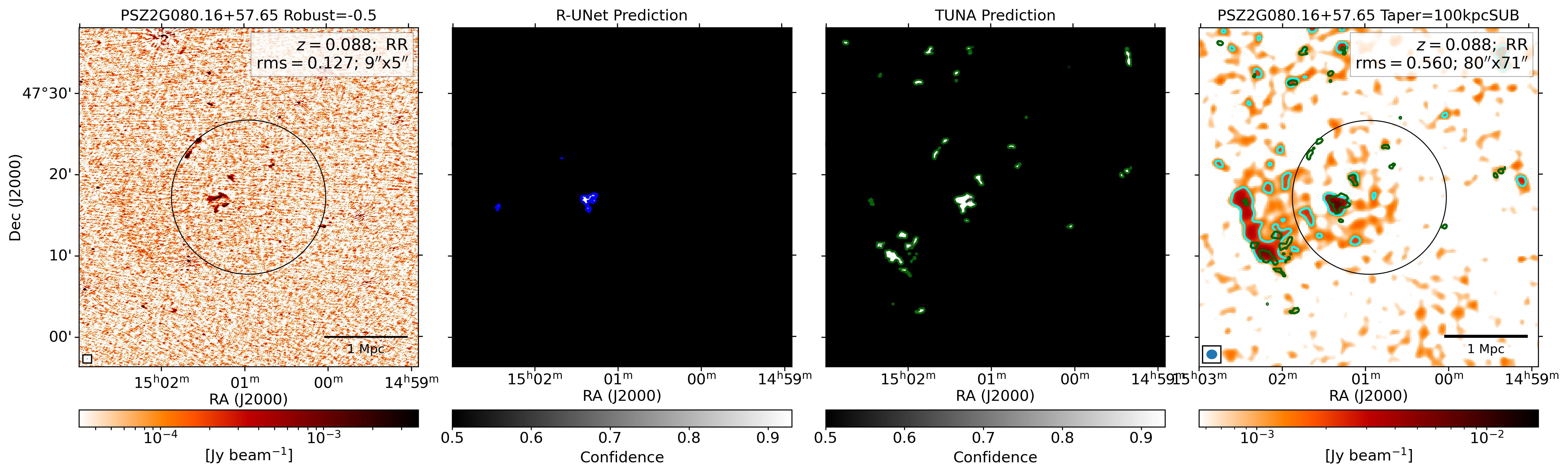}}\\
                \vspace{-1mm} 
                \subfloat{\raisebox{43ex}{\hspace{0ex}\footnotesize (d)}\includegraphics[width=1.1\textwidth]{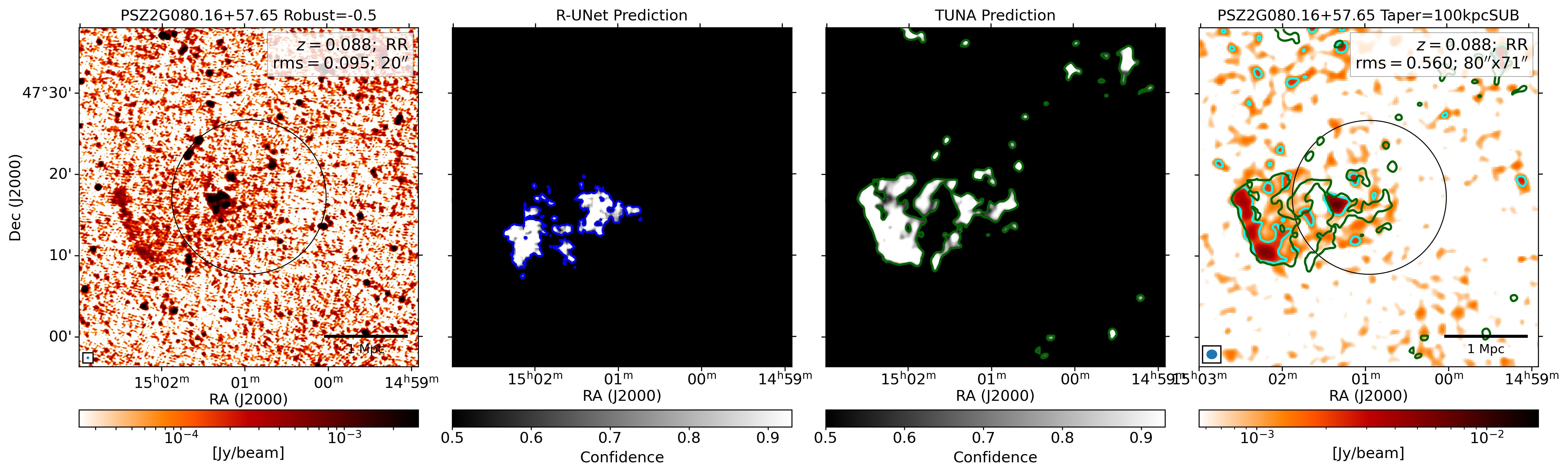}}
           \end{minipage}
           
        }
    }
   \caption{Left panel: radio images used as input for the networks. Central panels: prediction from the networks at a confidence level $\geq$ 0.5 with blue and green contours representing the confidence level = 0.5 for the R-UNet and TUNA networks, respectively. Right panel: 100 kpc tapered source-subtracted image with 3$\sigma$ level cyan contour and TUNA prediction green contour superimposed. The colour scale in the left panel is logarithmically stretched from 0.25 to 30$\sigma$; in the right panel, from 1 to 30$\sigma$. The redshift ($z$), image noise (rms, in mJy/beam units), classification, and image resolution of each cluster are reported in the top-right corner. The circle denotes the $R_{500}$ cluster radius.}    
    \label{fig:psz2}
\end{figure*}

The LoTSS-DR2/PSZ2 clusters have a redshift between $0.016 < z < 0.9$ and are classified as radio halos (RH), relics (RR), uncertain sources (U), candidate radio halo or candidate radio relic (cRH or cRR) in the absence of X-ray data. No diffuse emission (NDE) classifies objects that do not exhibit diffuse emission, though they may display lobes or tails from AGN in the field.  Not applicable (N/A) refers to objects that cannot be properly classified due to inadequate data quality. Clusters classified as N/A and those lacking redshift were excluded from analysis, resulting in a dataset of 246 objects. 
For each cluster, the images obtained using the Briggs weighting scheme  \citep{briggs95} with \texttt{Robust = -0.5} and no $uv$ taper, with a typical resolution of $\sim6''$, were used as input for TUNA and R-UNet. 

The performance of the network was evaluated by comparing its predictions with source-subtracted images, in which sensitivity to extended emission was enhanced through the removal of discrete sources from $uv$ data and subsequent re-imaging using a Gaussian $uv$ taper. We used images generated with \texttt{Robust = -0.5} and Gaussian $uv$ taper corresponding to 100 kpc at the cluster redshift, ensuring the detection of diffuse radio emission while accounting for resolution differences across redshifts (see Appendix \ref{sec:tables}). In this dataset, radio relics have been observed within the maximum distance of $2.2R_{500}$ \citep{Jones23}, hence $2 \times 2.2R_{500}$ cut-outs have been produced to perform our analysis.

\begin{figure*}
    \centering
    \makebox[\textwidth][c]{ %
        \resizebox{0.7\textwidth}{!}{ 
            \begin{minipage}{1\textwidth}
                \centering
                \subfloat{\raisebox{75ex}{\hspace{0ex} (a)}\includegraphics[width=1.0\textwidth]
                {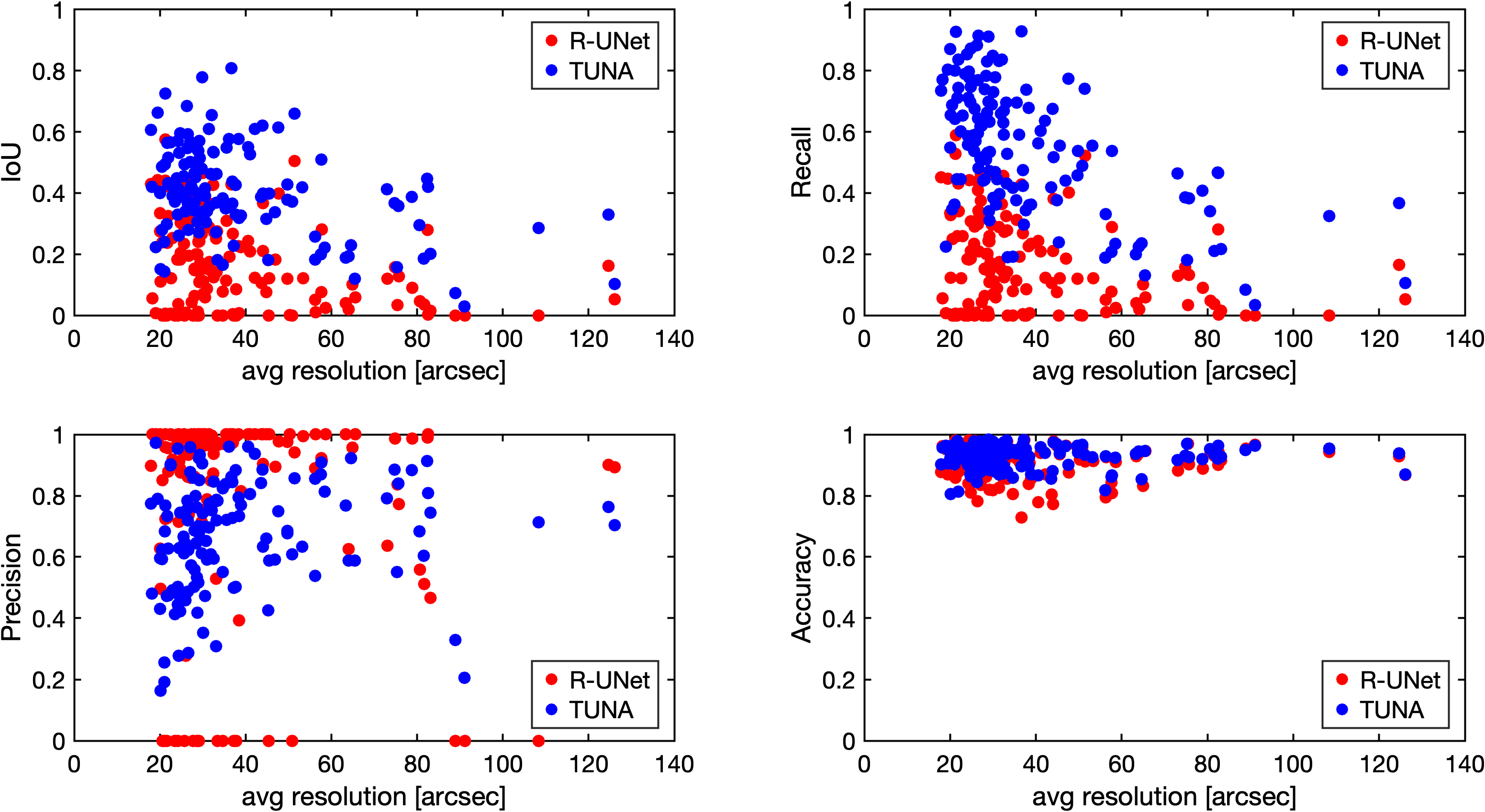}}\\
                \vspace{0mm} 
                \subfloat{\raisebox{75ex}{\hspace{0ex} (b)}\includegraphics[width=1.0\textwidth]{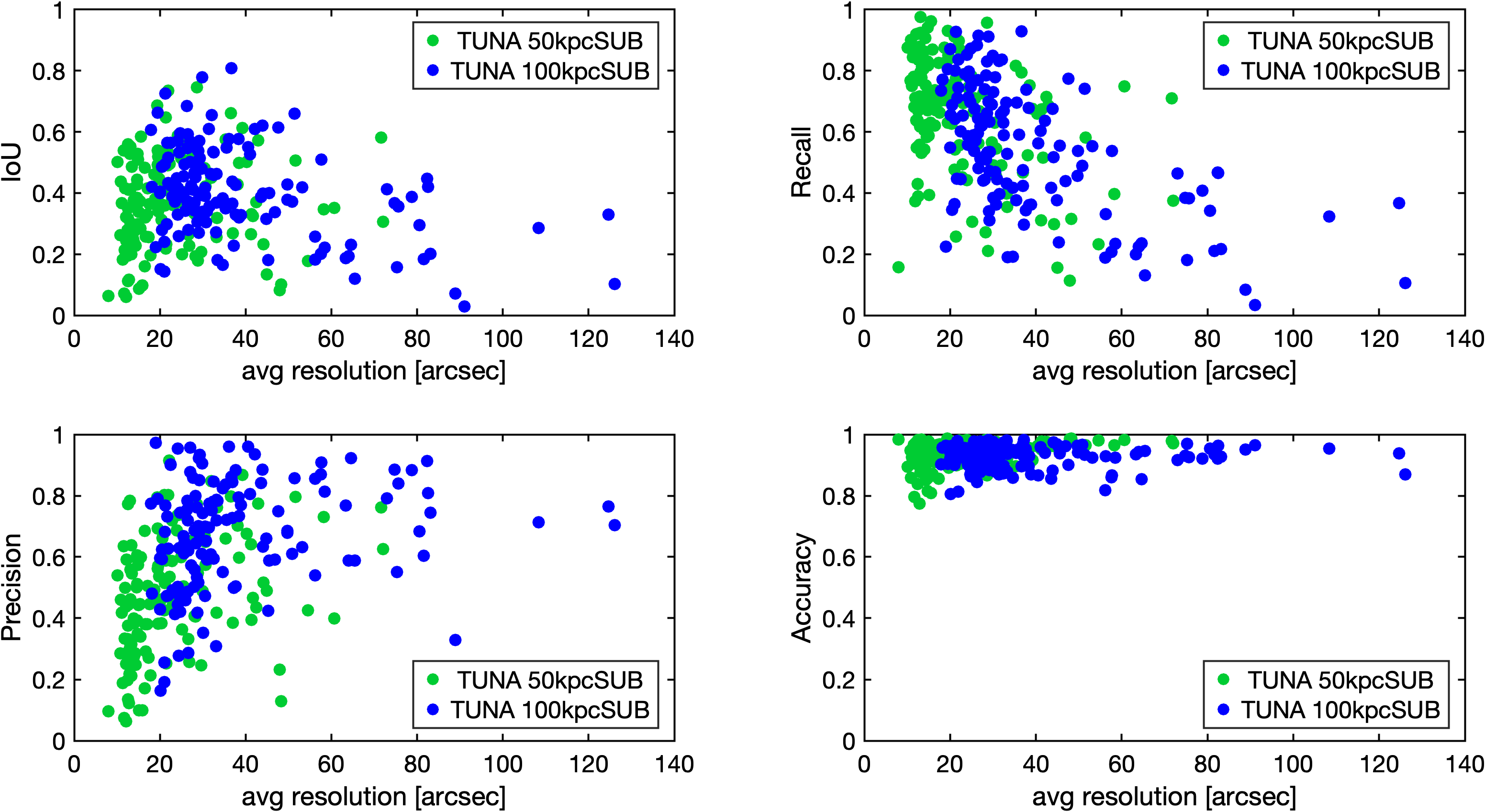}}\\
           \end{minipage}
           
        }
    }
   \caption{Panel (a): performance scores of TUNA and R-UNet as a function of the average resolution of the 100 kpc source-subtracted tapered images. Panel (b): performance metrics of TUNA as a function of the average resolution of the reference images, including 100 kpc and 50 kpc source-subtracted tapered images. The average resolution is calculated as $\left(B_{\rm maj} \times B_{\rm min}\right)^{1/2}$ where $B_{\rm maj}$ and $B_{\rm min}$ represent the major and minor axes of the beam, respectively.}  
    \label{fig:network_metric}
\end{figure*}

Panels (a), (b), and (c) of Fig. \ref{fig:psz2} show examples of the results generated by the networks, along with the input and reference images. TUNA accurately reproduces the morphology of the elongated radio halo in the cluster PSZ2G023.17+86.71, as shown in the 100 kpc source-subtracted tapered image. R-UNet only detects  the brightest areas of the radio halo. The PSZ2G166.62+42.13 cluster hosts a double radio relic, one prominent in the North-West and the other symmetrically positioned. Two additional fainter and smaller relics in the East and North-East directions have been identified. Additionally, a radio halo with low surface brightness is observed \citep{2024ApJ...966...38R}. R-UNet detects the two main relics and the brightest region of the radio halo, whereas TUNA reproduces precisely all the complex structures of the system. The PSZ2G080.16+57.65 is a low surface brightness giant radio relic-candidate radio halo system \citep{2021MNRAS.506..659C}. Both networks detect the radio halo, but only TUNA can partially identify the relic. Panel (d) shows that when using the low-resolution (20$\arcsec$) image \citep{shimwell2022lofar} as input, both networks successfully detect the diffuse emission associated with the radio halo and relic. However, TUNA identifies a larger region of diffuse emission and better traces the shape of the relic. This larger emission detected by TUNA is supported by observations in \cite{2021MNRAS.506..659C} using a $90''$ source-subtracted tapered image.

The masks generated by TUNA effectively replicate the size and morphology of the radio sources, identifying diffuse emission that becomes detectable only at much lower resolution than the input data and after subtraction of the embedded sources. This is confirmed by examining the entire image set from the catalogue LoTSS-DR2/PSZ2 (available on-line, see the Data Availability Section).  The network can remove the artifacts produced by the imaging process and exhibit the ability to detect a large number of radio galaxies within the field. 
  
The effectiveness of segmentation has been assessed using the four metrics defined in Section \ref{sec:optimization}: Recall, Precision, Intersection over Union (IoU), and Accuracy, calculated by comparing the inferred mask to the ground truth. Due to the unavailability of the true surface brightness distribution on the sky, the analysis used a mask derived from pixels with surface brightness above 3$\sigma$ in subtracted tapered images as a proxy for ground truth. This approach is considered a conservative approximation aimed at minimizing residual noise.
We considered the pixels inside a circular area of radius $R_c$, centered on the cluster's center, where $R_c = 2.2R_{500}$ for relics and $R_c = 1.1R_{500}$ for all other scenarios. The results obtained for each cluster are presented in Appendix \ref{sec:tables}.

Panel (a) of Fig. \ref{fig:network_metric} shows the relationship between the evaluation scores with the average resolution of the 100kpc source-subtracted tapered images (131 clusters, having neglected NDEs). The IoU and Recall metrics indicate that TUNA outperforms R-UNet in effectively capturing source morphology and ensuring detection completeness. Although R-UNet exhibits higher precision than TUNA, this is attributed to sub-segmentation, where only smaller portions of sources are identified, leading to fragmented detections that fail to capture the full structure of the sources. The accuracy, significantly affected by the large fraction of non-emitting regions, is close to unity.

Panel (b) of Fig. \ref{fig:network_metric} displays the evaluation metrics of TUNA, using both the 100kpc and 50kpc source-subtracted tapered images from \cite{botteon2022planck} as reference for the ground truth masks. Two separate taperings allows the assessment of TUNA's performance at different resolutions, pinpointing the specific resolution at which the network's segmentation mask is comparable with diffuse emission in lower-resolution images. 
The results indicate that TUNA achieves the best match within the resolution range of 20$''$ to 40$''$. In this range, IoU is higher, and Recall and Precision exhibit their best balance, with precision dropping at higher resolution and recall diminishing at lower resolution. The decreased precision is due to the smaller size of the sources with respect to the corresponding masks. This demonstrates TUNA's ability to identify faint diffuse emissions at a resolution 4-6 times lower than that of the input data ($\sim6''$).

The average evaluation metrics within the optimal range of $20''$ - $40''$ (see Table \ref{tab:metrics_obs}) confirm that TUNA outperforms R-UNet, with IoU more than twice as high and Recall more than three times greater. R-UNet has slightly higher Precision and both networks exhibit consistent Accuracy.

\begin{figure}
\begin{center}
	\includegraphics[width=0.7\columnwidth]{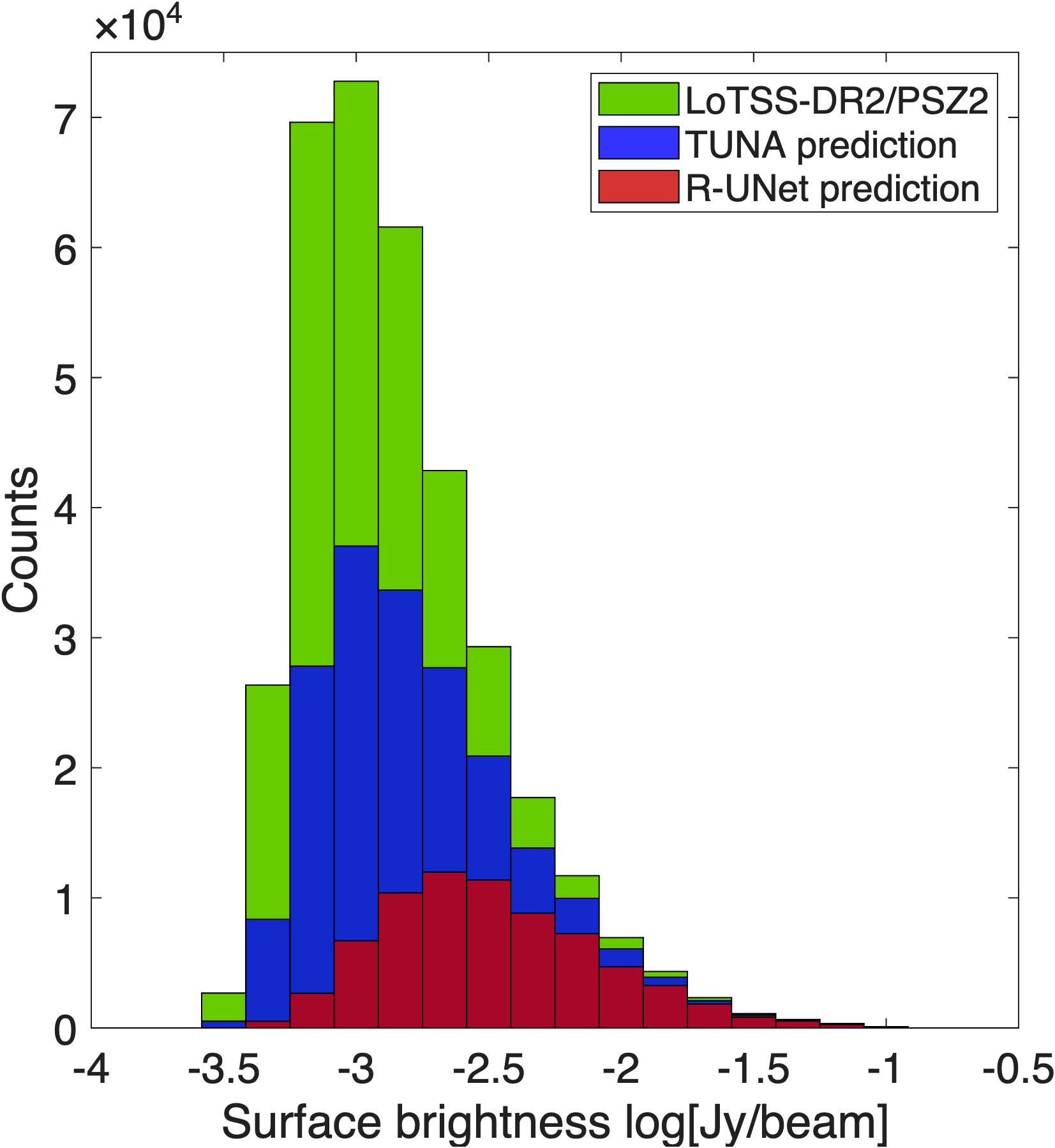}
    \caption{Surface brightness distribution for the LoTSS-DR2/PSZ2 observational catalogue considering pixels above $3\sigma$ (left), accounting for the pixels identified by TUNA and R-UNet}
    \label{fig:histoobs}
\end{center}
\end{figure}

Fig. \ref{fig:histoobs} shows the surface brightness for pixels above $3\sigma$, alongside the surface brightness identified by TUNA and R-UNet predictions that overlap this threshold mask. Above $10^{-2}\ {\rm Jy/beam}$, both networks detect the majority of source pixels. TUNA continues to identify a significant fraction (over 50\%) of the signal above $3\sigma$ extending down to $10^{-3}\ {\rm Jy/beam}$, the missing detections primarily due to the presence in the images of noise and artifacts exceeding $3\sigma$. Below $10^{-3}\ {\rm Jy/beam}$, TUNA keeps recognizing a significant fraction of emitting pixels. 

The capability of TUNA in identifying diffuse emission at resolutions 4-6 times lower than input data was also evaluated using low-resolution $20''$ images, as shown in panel (d) of Fig. \ref{fig:psz2}. The analysis compared masks generated using $20''$ images with those from 100kpc source-subtracted tapered images with resolution $\ge 70''$ (15 clusters in the entire dataset, available as on-line material, see Section \ref{sec:data_availability}). While the sample size  limits statistical reliability, the results suggest that TUNA can detect diffuse emission observable at resolutions between $60''$ and $120''$ (see also Section \ref{sec:megahalos}) .

\begin{table}
	\centering
	\caption{Performance metrics for TUNA and R-UNet calculated for the positive class (diffuse emission). The second and third columns present scores for the LoTSS-DR2/PSZ2 dataset.}
	\label{tab:metrics_obs}
	\begin{tabular}{lcc} 
		\hline
		Metric & TUNA & R-UNet \\
		\hline
        Recall & $0.61\pm 0.18$ & $0.20\pm 0.16$\\
		Precision & $0.65\pm 0.19$ & $0.77\pm 0.35$\\
        IoU & $0.43\pm 0.13$ & $0.19\pm 0.15$\\
        Accuracy & $0.93\pm 0.04$ & $0.92\pm 0.05$\\
		\hline
	\end{tabular}
\end{table}

\subsection{The A399-A401 Ridge and the Abell 1758 Bridge}

\begin{figure*}
\begin{center}
	\includegraphics[width=0.8\textwidth]{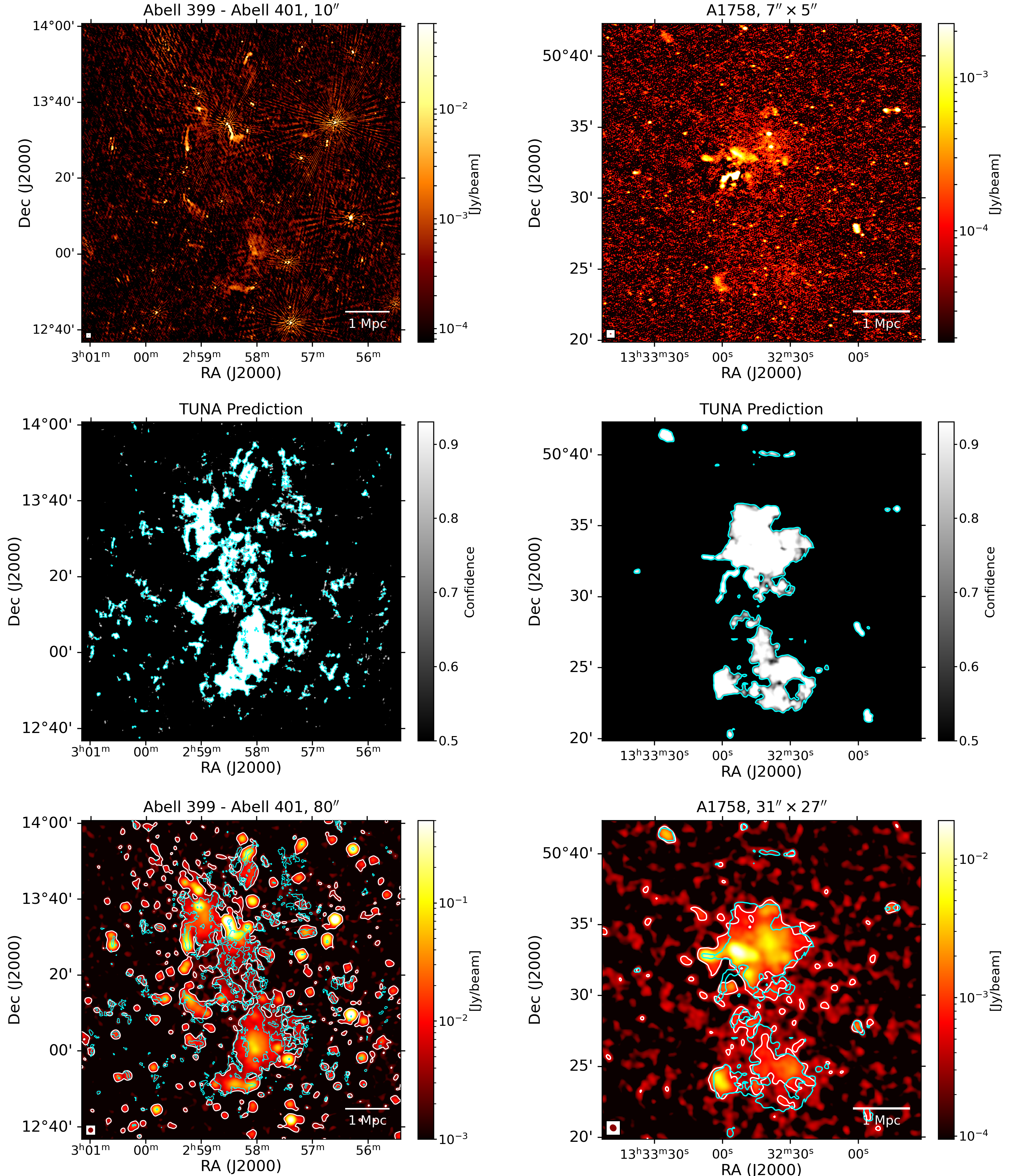}
    \caption{On the left, the A399-A401 ridge; on the right, the Abell 1758 bridge. Top panels: radio images used as input for TUNA. Central panels: TUNA predictions at 0.5 confidence level; cyan contours show the 0.9 confidence level for A399-A401 and 0.5 for A1758. Bottom panels: low resolution reference images with $3\sigma$ contours (white) and confidence contours (cyan) superimposed. The image resolution is shown on the top of each panel.}
    \label{fig:ridges}
\end{center}
\end{figure*}

Several recent low-frequency observations have shown the presence of diffuse radio emission between interacting cluster pairs in a pre-merger phase \citep[][]{govoni2019radio, botteon2020giant}. 

\begin{figure*}
\begin{center}
	\includegraphics[width=0.7\textwidth]{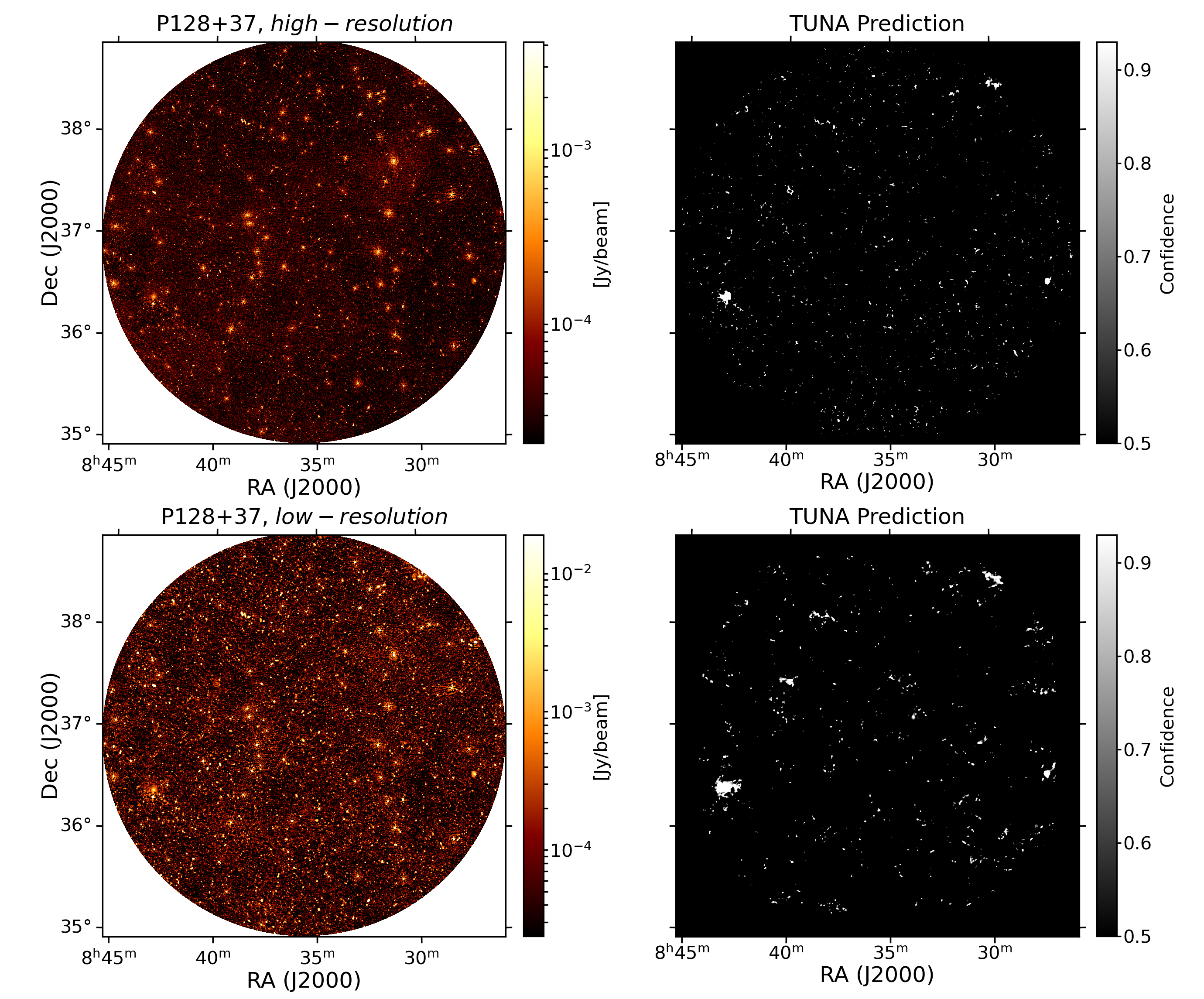}
    \caption{The LoTSS-DR2 pointing P128+37 enclosing the cluster Abell 697 (one of the four megahalos). The top row shows the high resolution ($6''$) radio observation (left panel) and the inferred TUNA mask (right panel). The bottom row shows the low resolution ($20''$) radio observation (left panel) and the inferred TUNA mask (right panel)}
    \label{fig:pointings}
\end{center}
\end{figure*}

\cite{govoni2019radio} observed a $\sim3~\rm{Mpc}$ ridge of radio emission connecting the pre-merging system made by Abell 399 and Abell 401 (A399-A401). The ridge was observed with the LOFAR high band antenna (HBA) at a central frequency 140 MHz. We use the $10''$ resolution image with rms sensitivity of about $\sigma= 0.3\; {\rm mJy/beam}$ as input for TUNA. Fig. \ref{fig:ridges}, left column, shows the comparison of the predicted mask with the 3$\sigma$ contour of the LOFAR observation at a resolution of $80''$ with sensitivity $\sigma= 1\; {\rm mJy/beam}$.  The confidence level for TUNA predictions is set to 0.9 to ensure reliability when the input resolution is lower than that used for training. Here, the input image has $10''$ resolution, while the network was trained on $6''$ resolution data.

\cite{botteon2020giant}, using LOFAR observations at 144 MHz, claimed the presence of a $\sim \rm 2~Mpc$ radio bridge connecting the pre-merging galaxy clusters A1758N and A1758S. Fig. \ref{fig:ridges}, right column, shows the mask predicted by TUNA. The image at resolution $\sim6''$  with sensitivity  $\sigma= 0.075\; {\rm mJy/beam}$ of the cluster PSZ2G107.10+65.32 from the LoTSS-DR2/PSZ2 dataset has been  used as input, whereas the 100 kpc tapered source-subtracted image at resolution $\sim30''$ with sensitivity  $\sigma= 0.19\; {\rm mJy/beam}$ has been used as comparison.

TUNA shows the capability to effectively probe low surface brightness diffuse emission from radio bridges, underscoring its crucial role in detecting rare and elusive sources.

\subsection{The Megahalos}
\label{sec:megahalos}

Based on LOFAR observations, \cite{cuciti2022galaxy} reported the existence of a new class of large diffuse radio sources that extend over 2–3 Mpc in four galaxy clusters, dubbed as radio megahalos. The volume of the megahalos is almost 30 times larger than that of radio halos and their emissivity is about 20 times lower. We also used TUNA to detect these low surface brightness sources. The input data for Abell 665, Abell 697 and Abell 2218 have been cutout from the $20''$ resolution pointings provided by the LoTSS-DR2 survey \citep{Shimwell22} (see Fig. \ref{fig:pointings}). The ZwCl 0634.1+4750 cutout at $30''$ resolution has been obtained with the same data reduction and imaging procedures used for the survey \citep{cuciti2022galaxy}.

\begin{figure*}
\begin{center}
	\includegraphics[width=0.65\textwidth]{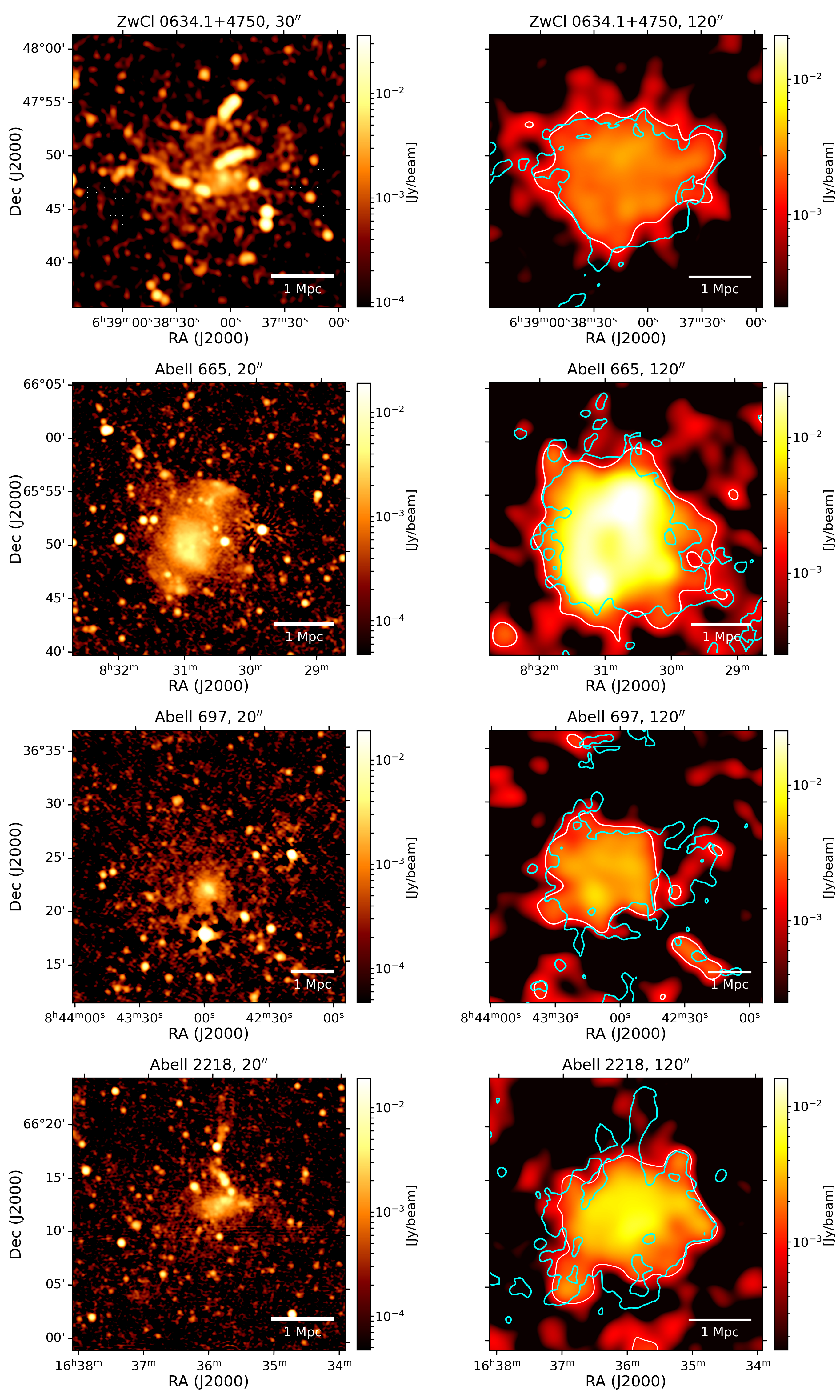}
    \caption{The four megahalos ZwCl 0634.1+4750, Abell 665, Abell 697 and Abell 2218. Left column: LOFAR 144 MHz radio images used as input for TUNA. Right column: low resolution, source subtracted images with $3\sigma$ contours (white) and TUNA confidence contours (cyan) superimposed. The confidence level is at 0.9 for ZwCl 0634.1+4750 and 0.5 for the remaining cases. The image resolution is shown on the top of each panel. The colour scale in the left panel uses a logarithmic stretch from 0.5 to 200$\sigma$, while the right panel uses 0.5 to 50$\sigma$. }
    \label{fig:megahalo}
\end{center}
\end{figure*}

For the megahalos, TUNA trained at $20''$ was used. The results are presented in Fig. \ref{fig:megahalo}, where confidence contours are overlaid on the $120''$ resolution images. The network accurately captures both their size and morphology, confirming once more its ability to detect diffuse emissions at significantly lower resolutions than the input data.

\section{Conclusions}
\label{sec:conclusions}

Artificial Intelligence offers novel solutions to achieve precise, automated, and rapid segmentation of astrophysical radio interferometric data. This paper demonstrates how Vision Transformers, integrated into the TransUNet architecture, enable new capabilities in radio astronomy image analysis, allowing for the detection of faint, diffuse radio sources that were  inaccessible with state-of-the-art CNNs. We introduced TUNA, a deep learning model based on Vision Transformers, trained on cosmological simulations and applied without retraining to real observational data, thereby directly evaluating its ability to generalize across domains. Our main achievements can be summarized as follows.

\begin{itemize}

\item 
TUNA detects extended radio sources at the sensitivity limits of the LOFAR HBA, without requiring manual source subtraction or low-resolution re-imaging. It can recover diffuse emission at scales equivalent to images reprocessed 4–6 times coarser than the original, while preserving morphological detail. Notable examples include the A399–A401 ridge and A1758 bridge, detected directly from high-resolution input, and all four known megahalos, originally observed in low-resolution images produced via substantial UV tapering. 
\smallskip

\item
TUNA enables fully automated data processing, requiring no additional human supervision. The network can exploit HPC solutions for both training and inference, which significantly reduces the time to solution. For instance, The LoTSS-DR2 pointing P128+37, shown in Fig. \ref{fig:pointings}, is processed in about 193 sec. at $6''$ ($9528\times 9528$ pixel) and in about 25 sec. at $20''$ ($3176\times 3176$ pixel), on Ampere100 NVIDIA GPUs of the Leonardo system at CINECA (see Section \ref{sec:optimization} for details). For comparison, the typical time needed to reduce one dataset to the cluster field, subtract discrete sources and produce a low resolution image could require up to one day.
\smallskip

\item
TUNA is highly effective in processing images with substantial noise and artefacts, recognising only diffused sources. However, when lower resolution images are used as input, the risk of confusion noise increases and blended point sources may be interpreted by the network as diffused objects. The most effective strategy to mitigate this issue is to avoid reprocessing the data at coarser resolution and instead leverage the network’s ability to detect extended emission the native telescope resolution, thereby preserving source morphology and reducing misclassification.
\smallskip

\item 
The network succeeds at generalising from simulated data to observational data.  Moreover, it generalize to diverse source types not present in the training set, including active galactic nuclei (AGN) and their associated jets. This capacity opens up new opportunities for blind source detection and classification.

\smallskip

\end{itemize}

The methodology established with TUNA is undergoing further extension. The simulated training set is augmented by using AGN and point sources obtained from actual observations. This will provide a more accurate and thorough identification of radio galaxies and compact objects, distinguishing them from other extended sources. Future developments also include expanding its applicability across frequencies and instruments, and enhancing its output to estimate physical properties such as total flux densities alongside segmentation.

\section*{Acknowledgements}

This paper is supported by the Fondazione ICSC, Spoke 3 Astrophysics and Cosmos Observations. National Recovery and Resilience Plan (Piano Nazionale di Ripresa e Resilienza, PNRR) Project ID CN\_00000013 "Italian Research Center for High-Performance Computing, Big Data and Quantum Computing" funded by MUR Missione 4 Componente 2 Investimento 1.4: Potenziamento strutture di ricerca e creazione di "campioni nazionali di R\&S (M4C2-19)" - Next Generation EU (NGEU), and it's also supported by (Programma Operativo Nazionale, PON), ``Tematiche di Ricerca Green e dell'Innovazione". We acknowledge the CINECA award under the ISCRA initiative, for the availability of high performance computing resources and support. The training and inference run this work is based on, have been produced on the Leonardo Supercomputer at CINECA (Bologna, Italy) in the framework of the ISCRA programme project IscrC\_ViTBuild. 
We also acknowledge the usage of online storage tools kindly provided by the INAF Astronomical Archive (IA2) initiative (http://www.ia2.inaf.it). 
We thank Andrea Botteon for the set up of the LoTSS-DR2/PSZ2 archive and Chiara Stuardi, B\"arbel Koribalski and Gabriele Giovannini for the insightful discussions.

\section*{Data Availability}
\label{sec:data_availability}

\noindent Results from the TUNA processing of LoTTS-DR2/PSZ2 data are publicly available at:

$\bullet$ https://owncloud.ia2.inaf.it/index.php/s/NtOMLBTXgJS2kZB

\smallskip

\noindent The Sky and Clean images produced and adopted for the training and testing of the network are publicly available in FITS format at: 

$\bullet$ https://owncloud.ia2.inaf.it/index.php/s/IbFPlCCcPUresrr 
\smallskip

\noindent The LOFAR radio images are publicly available at:

$\bullet$ https://lofar-surveys.org/index.html


\bibliographystyle{mnras}
\bibliography{bibliography}

\begin{thebibliography}{}
\makeatletter
\relax
\def\mn@urlcharsother{\let\do\@makeother \do\$\do\&\do\#\do\^\do\_\do\%\do\~}
\def\mn@doi{\begingroup\mn@urlcharsother \@ifnextchar [ {\mn@doi@}
  {\mn@doi@[]}}
\def\mn@doi@[#1]#2{\def\@tempa{#1}\ifx\@tempa\@empty \href
  {http://dx.doi.org/#2} {doi:#2}\else \href {http://dx.doi.org/#2} {#1}\fi
  \endgroup}
\def\mn@eprint#1#2{\mn@eprint@#1:#2::\@nil}
\def\mn@eprint@arXiv#1{\href {http://arxiv.org/abs/#1} {{\tt arXiv:#1}}}
\def\mn@eprint@dblp#1{\href {http://dblp.uni-trier.de/rec/bibtex/#1.xml}
  {dblp:#1}}
\def\mn@eprint@#1:#2:#3:#4\@nil{\def\@tempa {#1}\def\@tempb {#2}\def\@tempc
  {#3}\ifx \@tempc \@empty \let \@tempc \@tempb \let \@tempb \@tempa \fi \ifx
  \@tempb \@empty \def\@tempb {arXiv}\fi \@ifundefined
  {mn@eprint@\@tempb}{\@tempb:\@tempc}{\expandafter \expandafter \csname
  mn@eprint@\@tempb\endcsname \expandafter{\@tempc}}}

\bibitem[\protect\citeauthoryear{{Astropy Collaboration} et~al.,}{{Astropy
  Collaboration} et~al.}{2013}]{astropy:2013}
{Astropy Collaboration} et~al., 2013, \mn@doi [\aap]
  {10.1051/0004-6361/201322068}, \href
  {http://adsabs.harvard.edu/abs/2013A%26A...558A..33A} {558, A33}

\bibitem[\protect\citeauthoryear{Botteon et~al.,}{Botteon
  et~al.}{2020}]{botteon2020giant}
Botteon A.,  et~al., 2020, Monthly Notices of the Royal Astronomical Society:
  Letters, 499, L11

\bibitem[\protect\citeauthoryear{Botteon et~al.,}{Botteon
  et~al.}{2022}]{botteon2022planck}
Botteon A.,  et~al., 2022, Astronomy \& Astrophysics, 660, A78

\bibitem[\protect\citeauthoryear{{Briggs}}{{Briggs}}{1995}]{briggs95}
{Briggs} D.~S.,  1995, in American Astronomical Society Meeting Abstracts. p.
  112.02

\bibitem[\protect\citeauthoryear{{Brown}}{{Brown}}{2011}]{2011JApA...32..577B}
{Brown} S.~D.,  2011, \mn@doi [Journal of Astrophysics and Astronomy]
  {10.1007/s12036-011-9114-4}, \href
  {http://ads.ari.uni-heidelberg.de/abs/2011JApA...32..577B} {32, 577}

\bibitem[\protect\citeauthoryear{{Brunetti}}{{Brunetti}}{2011}]{Brunetti11b}
{Brunetti} G.,  2011, \mn@doi [Journal of Astrophysics and Astronomy]
  {10.1007/s12036-011-9103-7}, \href
  {https://ui.adsabs.harvard.edu/abs/2011JApA...32..437B} {32, 437}

\bibitem[\protect\citeauthoryear{Brunetti \& Jones}{Brunetti \&
  Jones}{2014}]{brunetti2014cosmic}
Brunetti G.,  Jones T.~W.,  2014, International Journal of Modern Physics D,
  23, 1430007

\bibitem[\protect\citeauthoryear{{Brunetti} \& {Vazza}}{{Brunetti} \&
  {Vazza}}{2020}]{brunetti20}
{Brunetti} G.,  {Vazza} F.,  2020, \mn@doi [\prl]
  {10.1103/PhysRevLett.124.051101}, \href
  {https://ui.adsabs.harvard.edu/abs/2020PhRvL.124e1101B} {124, 051101}

\bibitem[\protect\citeauthoryear{{Bryan} et~al.,}{{Bryan}
  et~al.}{2014}]{enzo14}
{Bryan} G.~L.,  et~al., 2014, \mn@doi [\apjs] {10.1088/0067-0049/211/2/19},
  \href {http://adsabs.harvard.edu/abs/2014ApJS..211...19B} {211, 19}

\bibitem[\protect\citeauthoryear{{Cavanagh}, {Bekki}  \& {Groves}}{{Cavanagh}
  et~al.}{2021}]{2021MNRAS.506..659C}
{Cavanagh} M.~K.,  {Bekki} K.,   {Groves} B.~A.,  2021, \mn@doi [Monthly
  Notices of the Royal Astronomical Society] {10.1093/mnras/stab1552}, \href
  {https://ui.adsabs.harvard.edu/abs/2021MNRAS.506..659C} {506, 659}

\bibitem[\protect\citeauthoryear{Chen et~al.,}{Chen
  et~al.}{2021}]{chen2021transunet}
Chen J.,  et~al., 2021, arXiv preprint arXiv:2102.04306

\bibitem[\protect\citeauthoryear{Cuciti et~al.,}{Cuciti
  et~al.}{2022}]{cuciti2022galaxy}
Cuciti V.,  et~al., 2022, Nature, 609, 911

\bibitem[\protect\citeauthoryear{{DeepSeek-AI} et~al.,}{{DeepSeek-AI}
  et~al.}{2024}]{2024arXiv241219437D}
{DeepSeek-AI} et~al., 2024, \mn@doi [arXiv e-prints]
  {10.48550/arXiv.2412.19437}, \href
  {https://ui.adsabs.harvard.edu/abs/2024arXiv241219437D} {p. arXiv:2412.19437}

\bibitem[\protect\citeauthoryear{Devlin, Chang, Lee  \& Toutanova}{Devlin
  et~al.}{2019}]{devlin-etal-2019-bert}
Devlin J.,  Chang M.-W.,  Lee K.,   Toutanova K.,  2019, in Burstein J.,  Doran
  C.,   Solorio T.,  eds, Proceedings of the 2019 Conference of the North
  {A}merican Chapter of the Association for Computational Linguistics: Human
  Language Technologies, Volume 1 (Long and Short Papers). Association for
  Computational Linguistics, Minneapolis, Minnesota, pp 4171--4186,
  \mn@doi{10.18653/v1/N19-1423}, \url {https://aclanthology.org/N19-1423/}

\bibitem[\protect\citeauthoryear{Dosovitskiy}{Dosovitskiy}{2020}]{dosovitskiy2020image}
Dosovitskiy A.,  2020, arXiv preprint arXiv:2010.11929

\bibitem[\protect\citeauthoryear{Dosovitskiy et~al.,}{Dosovitskiy
  et~al.}{2021}]{dosovitskiy2021an}
Dosovitskiy A.,  et~al., 2021, in International Conference on Learning
  Representations. \url {https://openreview.net/forum?id=YicbFdNTTy}

\bibitem[\protect\citeauthoryear{Feretti, Giovannini, Govoni  \&
  Murgia}{Feretti et~al.}{2012}]{feretti2012clusters}
Feretti L.,  Giovannini G.,  Govoni F.,   Murgia M.,  2012, The Astronomy and
  Astrophysics Review, 20, 1

\bibitem[\protect\citeauthoryear{{Gemini Team} et~al.,}{{Gemini Team}
  et~al.}{2023}]{2023arXiv231211805G}
{Gemini Team} et~al., 2023, \mn@doi [arXiv e-prints]
  {10.48550/arXiv.2312.11805}, \href
  {https://ui.adsabs.harvard.edu/abs/2023arXiv231211805G} {p. arXiv:2312.11805}

\bibitem[\protect\citeauthoryear{{Gheller} \& {Vazza}}{{Gheller} \&
  {Vazza}}{2022}]{2022MNRAS.509..990G}
{Gheller} C.,  {Vazza} F.,  2022, \mn@doi [\mnras] {10.1093/mnras/stab3044},
  \href {https://ui.adsabs.harvard.edu/abs/2022MNRAS.509..990G} {509, 990}

\bibitem[\protect\citeauthoryear{{Gheller}, {Vazza}  \& {Bonafede}}{{Gheller}
  et~al.}{2018}]{2018MNRAS.480.3749G}
{Gheller} C.,  {Vazza} F.,   {Bonafede} A.,  2018, \mn@doi [\mnras]
  {10.1093/mnras/sty2102}, \href
  {https://ui.adsabs.harvard.edu/abs/2018MNRAS.480.3749G} {480, 3749}

\bibitem[\protect\citeauthoryear{Govoni et~al.,}{Govoni
  et~al.}{2019}]{govoni2019radio}
Govoni F.,  et~al., 2019, Science, 364, 981

\bibitem[\protect\citeauthoryear{He, Zhang, Ren  \& Sun}{He
  et~al.}{2016}]{he2016resnet}
He K.,  Zhang X.,  Ren S.,   Sun J.,  2016, in Proceedings of the IEEE
  conference on computer vision and pattern recognition. pp 770--778

\bibitem[\protect\citeauthoryear{{Hodgson}, {Vazza}, {Johnston-Hollitt}  \&
  {McKinley}}{{Hodgson} et~al.}{2021}]{2021PASA...38...47H}
{Hodgson} T.,  {Vazza} F.,  {Johnston-Hollitt} M.,   {McKinley} B.,  2021,
  \mn@doi [\pasa] {10.1017/pasa.2021.32}, \href
  {https://ui.adsabs.harvard.edu/abs/2021PASA...38...47H} {38, e047}

\bibitem[\protect\citeauthoryear{{Hoeft} \& {Br{\"u}ggen}}{{Hoeft} \&
  {Br{\"u}ggen}}{2007}]{hb07}
{Hoeft} M.,  {Br{\"u}ggen} M.,  2007, \mn@doi [\mnras]
  {10.1111/j.1365-2966.2006.11111.x}, \href
  {http://adsabs.harvard.edu/abs/2007MNRAS.375...77H} {375, 77}

\bibitem[\protect\citeauthoryear{Hotan et~al.,}{Hotan
  et~al.}{2021}]{hotan2021australian}
Hotan A.,  et~al., 2021, Publications of the Astronomical Society of Australia,
  38, e009

\bibitem[\protect\citeauthoryear{Jonas \& Team}{Jonas \&
  Team}{2016}]{jonas2016meerkat}
Jonas J.,  Team M.,  2016, MeerKAT Science: On the Pathway to the SKA, p.~1

\bibitem[\protect\citeauthoryear{{Jones} et~al.,}{{Jones}
  et~al.}{2023}]{Jones23}
{Jones} A.,  et~al., 2023, \mn@doi [\aap] {10.1051/0004-6361/202245102}, \href
  {https://ui.adsabs.harvard.edu/abs/2023A&A...680A..31J} {680, A31}

\bibitem[\protect\citeauthoryear{Kohavi}{Kohavi}{1995}]{kohavi1995study}
Kohavi R.,  1995, in Proceedings of the 14th International Joint Conference on
  Artificial Intelligence (IJCAI). pp 1137--1143

\bibitem[\protect\citeauthoryear{{Lee}, {Pillepich}, {ZuHone}, {Nelson}, {Jee},
  {Nagai}  \& {Finner}}{{Lee} et~al.}{2023}]{Lee24}
{Lee} W.,  {Pillepich} A.,  {ZuHone} J.,  {Nelson} D.,  {Jee} M.~J.,  {Nagai}
  D.,   {Finner} K.,  2023, \mn@doi [arXiv e-prints]
  {10.48550/arXiv.2311.06340}, \href
  {https://ui.adsabs.harvard.edu/abs/2023arXiv231106340L} {p. arXiv:2311.06340}

\bibitem[\protect\citeauthoryear{{Nishiwaki}, {Brunetti}, {Vazza}  \&
  {Gheller}}{{Nishiwaki} et~al.}{2024}]{Nishiwaki24}
{Nishiwaki} K.,  {Brunetti} G.,  {Vazza} F.,   {Gheller} C.,  2024, \mn@doi
  [\apj] {10.3847/1538-4357/ad11ce}, \href
  {https://ui.adsabs.harvard.edu/abs/2024ApJ...961...15N} {961, 15}

\bibitem[\protect\citeauthoryear{Offringa \& Smirnov}{Offringa \&
  Smirnov}{2017}]{offringa-wsclean-2017}
Offringa A.~R.,  Smirnov O.,  2017, \mn@doi [MNRAS] {10.1093/mnras/stx1547},
  471, 301

\bibitem[\protect\citeauthoryear{Offringa, McKinley, Hurley-Walker
  et~al.}{Offringa et~al.}{2014}]{offringa-wsclean-2014}
Offringa A.~R.,  McKinley B.,  Hurley-Walker  et~al., 2014, \mn@doi [MNRAS]
  {10.1093/mnras/stu1368}, 444, 606

\bibitem[\protect\citeauthoryear{{Pfrommer}, {Springel}, {En{\ss}lin}  \&
  {Jubelgas}}{{Pfrommer} et~al.}{2006}]{pf06}
{Pfrommer} C.,  {Springel} V.,  {En{\ss}lin} T.~A.,   {Jubelgas} M.,  2006,
  \mn@doi [\mnras] {10.1111/j.1365-2966.2005.09953.x}, \href
  {http://adsabs.harvard.edu/abs/2006MNRAS.367..113P} {367, 113}

\bibitem[\protect\citeauthoryear{{Pignataro} et~al.,}{{Pignataro}
  et~al.}{2024}]{2024A&A...691A..99P}
{Pignataro} G.~V.,  et~al., 2024, \mn@doi [\aap] {10.1051/0004-6361/202451529},
  \href {https://ui.adsabs.harvard.edu/abs/2024A&A...691A..99P} {691, A99}

\bibitem[\protect\citeauthoryear{Radford, Narasimhan, Salimans  \&
  Sutskever}{Radford et~al.}{2018}]{radford2018improving}
Radford A.,  Narasimhan K.,  Salimans T.,   Sutskever I.,  2018, OpenAI
  Technical Report

\bibitem[\protect\citeauthoryear{Radovic et~al.,}{Radovic
  et~al.}{2018}]{radovic2018machine}
Radovic A.,  et~al., 2018, Nature, 560, 41

\bibitem[\protect\citeauthoryear{{Rajpurohit} et~al.,}{{Rajpurohit}
  et~al.}{2024}]{2024ApJ...966...38R}
{Rajpurohit} K.,  et~al., 2024, \mn@doi [\apj] {10.3847/1538-4357/ad29fa},
  \href {https://ui.adsabs.harvard.edu/abs/2024ApJ...966...38R} {966, 38}

\bibitem[\protect\citeauthoryear{Rolnick et~al.,}{Rolnick
  et~al.}{2022}]{rolnick2022climate_change}
Rolnick D.,  et~al., 2022, ACM Computing Surveys (CSUR), 55, 1

\bibitem[\protect\citeauthoryear{Ronneberger, Fischer  \& Brox}{Ronneberger
  et~al.}{2015}]{ronneberger2015}
Ronneberger O.,  Fischer P.,   Brox T.,  2015, Medical Image Computing and
  Computer-Assisted Intervention – MICCAI 2015

\bibitem[\protect\citeauthoryear{{Ryu}, {Kang}, {Hallman}  \& {Jones}}{{Ryu}
  et~al.}{2003}]{ry03}
{Ryu} D.,  {Kang} H.,  {Hallman} E.,   {Jones} T.~W.,  2003, \mn@doi [\apj]
  {10.1086/376723}, \href {http://adsabs.harvard.edu/abs/2003ApJ...593..599R}
  {593, 599}

\bibitem[\protect\citeauthoryear{{Ryu}, {Kang}, {Cho}  \& {Das}}{{Ryu}
  et~al.}{2008}]{2008Sci...320..909R}
{Ryu} D.,  {Kang} H.,  {Cho} J.,   {Das} S.,  2008, \mn@doi [Science]
  {10.1126/science.1154923}, \href
  {https://ui.adsabs.harvard.edu/abs/2008Sci...320..909R} {320, 909}

\bibitem[\protect\citeauthoryear{{Sanvitale}, {Gheller}  \&
  {Bowman}}{{Sanvitale} et~al.}{2022}]{Sanvitale2022}
{Sanvitale} N.,  {Gheller} C.,   {Bowman} E.,  2022, \mn@doi [Granular Matter]
  {10.1007/s10035-022-01222-w}, 24

\bibitem[\protect\citeauthoryear{Shimwell et~al.,}{Shimwell
  et~al.}{2022a}]{shimwell2022lofar}
Shimwell T.,  et~al., 2022a, Astronomy \& astrophysics, 659, A1

\bibitem[\protect\citeauthoryear{{Shimwell} et~al.,}{{Shimwell}
  et~al.}{2022b}]{Shimwell22}
{Shimwell} T.~W.,  et~al., 2022b, \mn@doi [\aap] {10.1051/0004-6361/202142484},
  \href {https://ui.adsabs.harvard.edu/abs/2022A&A...659A...1S} {659, A1}

\bibitem[\protect\citeauthoryear{{Stuardi}, {Gheller}, {Vazza}  \&
  {Botteon}}{{Stuardi} et~al.}{2024}]{2024MNRAS.533.3194S}
{Stuardi} C.,  {Gheller} C.,  {Vazza} F.,   {Botteon} A.,  2024, \mn@doi
  [\mnras] {10.1093/mnras/stae2014}, \href
  {https://ui.adsabs.harvard.edu/abs/2024MNRAS.533.3194S} {533, 3194}

\bibitem[\protect\citeauthoryear{Tingay et~al.,}{Tingay
  et~al.}{2013}]{tingay2013murchison}
Tingay S.~J.,  et~al., 2013, Publications of the Astronomical Society of
  Australia, 30, e007

\bibitem[\protect\citeauthoryear{Vamathevan et~al.,}{Vamathevan
  et~al.}{2019}]{vamathevan2019applications}
Vamathevan J.,  et~al., 2019, Nature reviews Drug discovery, 18, 463

\bibitem[\protect\citeauthoryear{Van~Weeren, de Gasperin, Akamatsu,
  Br{\"u}ggen, Feretti, Kang, Stroe  \& Zandanel}{Van~Weeren
  et~al.}{2019}]{van2019diffuse}
Van~Weeren R.,  de Gasperin F.,  Akamatsu H.,  Br{\"u}ggen M.,  Feretti L.,
  Kang H.,  Stroe A.,   Zandanel F.,  2019, Space Science Reviews, 215, 1

\bibitem[\protect\citeauthoryear{Vaswani, Shazeer, Parmar, Uszkoreit, Jones,
  Gomez, Kaiser  \& Polosukhin}{Vaswani et~al.}{2017}]{vaswani2017attention}
Vaswani A.,  Shazeer N.,  Parmar N.,  Uszkoreit J.,  Jones L.,  Gomez A.~N.,
  Kaiser L.,   Polosukhin I.,  2017, arXiv preprint arXiv:1706.03762, 10,
  S0140525X16001837

\bibitem[\protect\citeauthoryear{{Vazza}, {Paoletti}, {Banfi}, {Finelli},
  {Gheller}, {O'Sullivan}  \& {Br{\"u}ggen}}{{Vazza}
  et~al.}{2021}]{2021MNRAS.500.5350V}
{Vazza} F.,  {Paoletti} D.,  {Banfi} S.,  {Finelli} F.,  {Gheller} C.,
  {O'Sullivan} S.~P.,   {Br{\"u}ggen} M.,  2021, \mn@doi [\mnras]
  {10.1093/mnras/staa3532}, \href
  {https://ui.adsabs.harvard.edu/abs/2021MNRAS.500.5350V} {500, 5350}

\bibitem[\protect\citeauthoryear{{Vazza}, {Gheller}, {Zanetti}, {Tsizh},
  {Carretti}, {Mtchedlidze}  \& {Brueggen}}{{Vazza}
  et~al.}{2025}]{2025arXiv250119041V}
{Vazza} F.,  {Gheller} C.,  {Zanetti} F.,  {Tsizh} M.,  {Carretti} E.,
  {Mtchedlidze} S.,   {Brueggen} M.,  2025, \mn@doi [arXiv e-prints]
  {10.48550/arXiv.2501.19041}, \href
  {https://ui.adsabs.harvard.edu/abs/2025arXiv250119041V} {p. arXiv:2501.19041}

\bibitem[\protect\citeauthoryear{{Wells}, {Greisen}  \& {Harten}}{{Wells}
  et~al.}{1981}]{1981A&AS...44..363W}
{Wells} D.~C.,  {Greisen} E.~W.,   {Harten} R.~H.,  1981, \aaps, \href
  {https://ui.adsabs.harvard.edu/abs/1981A&AS...44..363W} {44, 363}

\bibitem[\protect\citeauthoryear{Wittor}{Wittor}{2023}]{wittor2023cosmic}
Wittor D.,  2023, Universe, 9, 319

\bibitem[\protect\citeauthoryear{van Haarlem et~al.,}{van Haarlem
  et~al.}{2013}]{van2013lofar}
van Haarlem M.~P.,  et~al., 2013, Astronomy \& astrophysics, 556, A2

\makeatother
\end{thebibliography}

\appendix
\newpage

\section{Hyperparameters Optimization Analysis}
\label{sec:hyper}

Table A1 presents the results of the evaluation of the impact of different hyperparameter settings on model performance. We varied tile size, batch size, learning rate, and Class 1 weight, and report the resulting Intersection over Union (IoU), precision, recall, and an average score combining the three metrics. The best 15 models are presented. The first line of the Table shows our reference model. Changes in the hyperparameters result in small changes to the composite average score. Our tests indicates that a tile size of 512 and a learning rate between 0.001 and 0.005 yield optimal results. Learning rates smaller than 0.001 are absent from the initial 15 scores, due to lower performance. No clear trend was observed for batch size. Therefore, we adopted the value used in the original TransUnet implementation. 
The Class 1 weight $w_1$ is a multiplicative factor applied to the loss contribution of Class 1 pixels during training. It increases the contribution of underrepresented Class 1 pixels to the overall loss. In principle, this approach should improve model accuracy for the underrepresented class. In practice, however, it proved ineffective.

\begin{table*}
\centering
\caption{Performance metrics (IoU, Precision, Recall, and composite Average Score) of the network across various hyperparameter configurations. Each row corresponds to a different combination of tile size, batch size, learning rate, and Class 1 weight $w_1$. The average score is computed as the mean of IoU, precision, and recall to provide a single comparative value.}
\begin{tabular}{cccccccc} 
\hline
Tile size & Batch Size & Learning Rate & $w_1$ & IoU & Precision & Recall & Avg. Score\\ 
\hline
512 & 24 & 0.005 & 1 & 0.38 & 0.61 & 0.50 & 0.50\\ \hline
512 & 12 & 0.001 & 1 & 0.40	& 0.61 & 0.53 & 0.51\\ \hline
512 & 12  & 0.005 & 1 & 0.39& 0.58 & 0.52 & 0.50\\ \hline
512 & 12 & 0.01  & 1 & 0.38	& 0.60 & 0.51 & 0.50\\ \hline
512 & 24 & 0.001 & 1 & 0.39 & 0.58 & 0.49 & 0.49\\ \hline
224 & 48 & 0.001 & 1 & 0.36	& 0.64 & 0.47 & 0.49\\ \hline
224 & 12 & 0.005 & 1 & 0.36	& 0.60 & 0.48 & 0.48\\ \hline
224 & 24 & 0.001 & 2 & 0.36	& 0.61 & 0.47 & 0.48\\ \hline
512 & 24 & 0.01  & 1 & 0.37	& 0.53 & 0.54 & 0.47\\ \hline
224 & 48 & 0.01  & 1 & 0.35	& 0.64 & 0.43 & 0.47\\ \hline
224 & 24 & 0.01  & 1 & 0.35 & 0.58 & 0.48 & 0.47\\ \hline
224 & 24 & 0.001 & 1 & 0.35	& 0.57 & 0.49 & 0.47\\ \hline
224 & 24 & 0.005 & 1 & 0.34 & 0.61 & 0.45 & 0.47\\ \hline
224 & 12 & 0.01  & 1 & 0.35	& 0.61 & 0.45 & 0.47\\ \hline
224 & 24 & 0.001 & 10& 0.34 & 0.46 & 0.59 & 0.46\\ \hline
\end{tabular}
\end{table*}

\section{Supplementary Tables}
\label{sec:tables}

This appendix presents sample tables detailing the performance scores of TUNA (Table B1)  and R-UNet (Table B2) as applied to the LoTSS-DR2/PSZ2 survey clusters at an average resolution of 100 kpc, as well as the performance of TUNA on clusters at a resolution of 50 kpc (Table B3). The complete data tables, in tab-separated ASCII (TSV) format, are included as supplementary material. 

\begin{table*}
\centering
\caption{Performance scores of TUNA, considering the LoTSS-DR2/PSZ2 survey clusters (``Not Applicable" clusters excluded). The last column shows the average resolution of the 100kpc source-subtracted tapered images used for evaluating the network's predictions. The average resolution is calculated as $\left(B_{\rm maj} \times B_{\rm min}\right)^{1/2}$ where $B_{\rm maj}$ and $B_{\rm min}$ represent the major and minor axes of the beam, respectively. The full data table in ASCII format is available on-line as supplementary material.} 
\begin{tabular}{lccccccc} 
\hline
Cluster ID & Classification & Redshift & IoU & Precision & Recall & Accuracy & avg res 100kpcSUB \\ 
\hline
PSZ2G023.17+86.71 & RH & 0.306 & 0.481 & 0.534 & 0.830 & 0.908 & 29 \\ \hline
PSZ2G031.93+78.71 & RH, U & 0.072 & 0.447 & 0.913 & 0.467 & 0.921 & 82 \\ \hline
PSZ2G040.58+77.12 & RH & 0.075 & 0.295 & 0.684 & 0.341 & 0.953 & 81 \\ \hline
PSZ2G045.13+67.78 & NDE & 0.219 & 0.168 & 0.450 & 0.211 & 0.963 & 45 \\ \hline
PSZ2G045.87+57.70 & RH & 0.611 & 0.404 & 0.430 & 0.871 & 0.921 & 20 \\ \hline
PSZ2G046.88+56.48 & RH & 0.115 & 0.182 & 0.856 & 0.188 & 0.818 & 56 \\ \hline
PSZ2G048.10+57.16 & RH, RR & 0.078 & 0.387 & 0.884 & 0.408 & 0.920 & 79 \\ \hline
PSZ2G048.75+53.18 & NDE & 0.098 & 0.102 & 0.188 & 0.183 & 0.970 & 64 \\ \hline
PSZ2G049.18+65.05 & NDE & 0.234 & 0.377 & 0.605 & 0.500 & 0.990 & 32 \\ \hline
PSZ2G049.32+44.37 & RH & 0.097 & 0.188 & 0.768 & 0.199 & 0.936 & 63 \\ \hline
\end{tabular}
\end{table*}

\begin{table*}
\centering
\caption{Performance scores of R-UNet on the LoTSS-DR2/PSZ2 survey clusters ("Not Applicable" clusters excluded). The last column shows the average resolution of the 100kpc source-subtracted tapered images used for evaluating the network's predictions. The average resolution is calculated as $\left(B_{\rm maj} \times B_{\rm min}\right)^{1/2}$ where $B_{\rm maj}$ and $B_{\rm min}$ represent the major and minor axes of the beam, respectively. The full data table in ASCII format is available on-line as supplementary material.} 
\begin{tabular}{lccccccc} 
\hline
Cluster ID & Classification & Redshift & IoU & Precision & Recall & Accuracy & avg res 100kpcSUB \\ 
\hline
PSZ2G023.17+86.71 & RH & 0.306 & 0.199 & 0.976 & 0.200 & 0.917 & 29 \\ \hline
PSZ2G031.93+78.71 & RH, U & 0.072 & 0.279 & 0.989 & 0.280 & 0.901 & 82 \\ \hline
PSZ2G040.58+77.12 & RH & 0.075 & 0.047 & 0.558 & 0.048 & 0.943 & 81 \\ \hline
PSZ2G045.13+67.78 & NDE & 0.219 & 0.000 & 0.000 & 0.000 & 0.965 & 45 \\ \hline
PSZ2G045.87+57.70 & RH & 0.611 & 0.275 & 0.627 & 0.328 & 0.947 & 20 \\ \hline
PSZ2G046.88+56.48 & RH & 0.115 & 0.051 & 0.890 & 0.051 & 0.794 & 56 \\ \hline
PSZ2G048.10+57.16 & RH, RR & 0.078 & 0.090 & 0.987 & 0.090 & 0.888 & 79 \\ \hline
PSZ2G048.75+53.18 & NDE & 0.098 & 0.000 & 0.000 & 0.000 & 0.977 & 64 \\ \hline
PSZ2G049.18+65.05 & NDE & 0.234 & 0.000 & 0.000 & 0.000 & 0.988 & 32 \\ \hline
PSZ2G049.32+44.37 & RH & 0.097 & 0.039 & 1.000 & 0.039 & 0.928 & 63 \\ \hline
\end{tabular}
\end{table*}

\begin{table*}
\centering
\caption{Performance scores of TUNA considering the LoTSS-DR2/PSZ2 survey clusters ("Not Applicable" clusters excluded). The last column shows the average resolution of the 50kpc source-subtracted tapered images used for evaluating the predictions of the network. The average resolution is calculated as $\left(B_{\rm maj} \times B_{\rm min}\right)^{1/2}$ where $B_{\rm maj}$ and $B_{\rm min}$ represent the major and minor axes of the beam, respectively. The full data table in ASCII format is available on-line as supplementary material.} 
\begin{tabular}{lccccccc} 
\hline
Cluster ID & Classification & Redshift & IoU & Precision & Recall & Accuracy & avg res 50kpcSUB \\ 
\hline
PSZ2G023.17+86.71 & RH & 0.306 & 0.409 & 0.426 & 0.912 & 0.901 & 21 \\ \hline
PSZ2G031.93+78.71 & RH, U & 0.072 & 0.572 & 0.775 & 0.686 & 0.959 & 43 \\ \hline
PSZ2G040.58+77.12 & RH & 0.075 & 0.331 & 0.395 & 0.674 & 0.977 & 41 \\ \hline
PSZ2G045.13+67.78 & NDE & 0.219 & 0.227 & 0.342 & 0.402 & 0.981 & 19 \\ \hline
PSZ2G045.87+57.70 & RH & 0.611 & 0.245 & 0.250 & 0.920 & 0.904 & 12 \\ \hline
PSZ2G046.88+56.48 & RH & 0.115 & 0.324 & 0.489 & 0.491 & 0.950 & 30 \\ \hline
PSZ2G048.10+57.16 & RH, RR & 0.078 & 0.501 & 0.676 & 0.660 & 0.962 & 40 \\ \hline
PSZ2G048.75+53.18 & NDE & 0.098 & 0.123 & 0.160 & 0.348 & 0.979 & 34 \\ \hline
PSZ2G049.18+65.05 & NDE & 0.234 & 0.307 & 0.533 & 0.420 & 0.988 & 22 \\ \hline
PSZ2G049.32+44.37 & RH & 0.097 & 0.256 & 0.419 & 0.398 & 0.977 & 33 \\ \hline

\end{tabular}
\end{table*}

\end{document}